\documentclass[namedreferences]{solarphysics}
 \pdfoutput=1
 
\usepackage[optionalrh,solaromanenum,natbib]{spr-sola-addons} 
\usepackage{natbib}
\usepackage{epsfig}                     
\usepackage{graphicx}                    

\usepackage{courier}                    
\usepackage{amssymb}                    
\usepackage{color}                       

\usepackage{url}                         

\usepackage{pdflscape}

\newcommand{\aap}{    {\it Astron. Astrophys.}}

\newcommand{\apj}{    {\it Astrophys. J.}}
\newcommand{\apjl}{   {\it Astrophys. J. Lett.}}

\newcommand{\grl}{    {\it Geophys. Res. Lett.}}

\newcommand{\jgr}{    {\it J. Geophys. Res.}}

\newcommand{\nat}{    {\it Nature}}

\newcommand{\solphys}{{\it Solar Phys.}}
 
\newcommand{\ssr}{    {\it Space Sci. Rev.}}

\newcommand{\zap}{  {\it Z. Astrophys. }}

\newcommand{\icarus}{  {\it Icarus }}
\newcommand{\apjs}{  {\it ApJS }}

\begin{document}

\begin{article}

\begin{opening}

\title{Evolution and Consequences of Interacting CMEs of 2012 November 9-10 using STEREO/SECCHI and In Situ Observations}

\author{Wageesh~\surname{Mishra}$^{1}$\sep
        Nandita~\surname{Srivastava}$^{1}$\sep D.~\surname{Chakrabarty}$^{2}$   
       }

\runningauthor{Mishra et al.}
\runningtitle{Evolution and Consequences of Interacting CMEs}

\institute{$^{1}$ Udaipur Solar Observatory, Physical Research Laboratory, Udaipur, India -313001 \\
                     email: \url{wageesh@prl.res.in} email: \url{nandita@prl.res.in}
             }
						
\institute{$^{2}$ Space and Atmospheric Sciences Division, Physical Research Laboratory, Ahmedabad-380009, India\\
                    email: \url{dipu@prl.res.in} 
										}

\begin{abstract}
Understanding of the kinematic evolution of Coronal Mass Ejections (CMEs) in the heliosphere is important to estimate their arrival time at the Earth. It is found that kinematics of CMEs can change when they interact or collide with each other as they propagate in the heliosphere. In this paper, we analyze the collision and post-interaction characteristics of two Earth-directed CMEs, launched successively on 2012 November 9 and 10, using white light imaging observations from STEREO/SECCHI and in situ observations taken from WIND spacecraft. We tracked two density enhanced features associated with leading and trailing edge of November 9 CME and one density enhanced feature associated with leading edge of November 10 CME by constructing J-maps. We found that the leading edge of November 10 CME interacted with the trailing edge of November 9 CME. We also estimated the kinematics of these features of the CMEs and found a significant change in their dynamics after interaction. In in situ observations, we identified distinct structures associated with interacted CMEs and also noticed their heating and compression as signatures of CME-CME interaction. Our analysis shows an improvement in arrival time prediction of CMEs using their post-collision dynamics than using pre-collision dynamics. Estimating the true masses and speeds of these colliding CMEs, we investigated the nature of observed collision which is found to be close to perfectly inelastic. The investigation also places in perspective the geomagnetic consequences of the two CMEs and their interaction in terms of occurrence of geomagnetic storm and triggering of magnetospheric substorms.

\end{abstract}


\end{opening}

\section{Introduction}
\label{Intro}
Coronal Mass Ejections (CMEs) are huge magnetized plasma expulsions from the Sun into the heliosphere and are considered as primary drivers of geomagnetic storms and various other space weather effects \citep{Gosling1990,Echer2008}. It is therefore required to improve the physics and our understanding of evolution of CMEs in the solar wind, which is found to be dependent on heliospheric environment \citep{Kilpua2012, Vrsnak2013}. Various techniques based on stereoscopic observations which estimate the 3D kinematics of CMEs in the heliosphere have been developed \citep{Mierla2008,Thernisien2009,Thompson2009,Liu2010,Lugaz2010.apj,Davies2013}. Using these techniques, association of remote observations of CMEs with in situ observations has been carried out by \citet{Liu2010.722,Temmer2011,Kilpua2012,Mostl2012,Mishra2013,Mishra2014.apj}. Also, several models have been developed for predicting the arrival time of CMEs at 1 AU \citep{Gopalswamy2000,Vrsnak2013}. Sometimes, CMEs are observed to erupt in quick succession which under certain favorable initial conditions, can interact or merge with each other during their propagation in the heliosphere. In such a scenario, space weather prediction schemes may not be successful without taking into account the post-interaction CMEs characteristics. The compound streams (interaction of CME-CIR or CME-CME) were first inferred by \citet{Burlaga1987} using observations from Helios and ISEEE 3 spacecraft. Their study suggested that a large geomagnetic storm may be associated with compound streams. The observational evidence for merging of CMEs (or cannibalism) was first given by \citet{Gopalswamy2001apj} based on the long-wavelength radio and SOHO/LASCO observations. Thus, these interactions are of great importance from space weather point of view. Before the era of wide angle imaging far from the Sun, the understanding of involved physical mechanisms in CME-CME or CME-shock interaction was achieved mostly from magnetohydrodynamic (MHD) numerical simulations of the interaction of a shock wave with an magnetic cloud (MC) \citep{Vandas1997,Vandas2004,Xiong2006}, the interaction of two ejecta \citep{Gonzalez-Esparza2004,Lugaz2005,Wang2005}, and the interaction of two MCs \citep{Xiong2007,Xiong2009}. \citet{Wang2003} have shown that a forward shock can cause an intense southward magnetic field of long duration in the preceding magnetic cloud. Such modifications in the preceding cloud are important for space weather prediction. Therefore, events involving CME-CME and CME-shock interactions are important candidates to investigate their kinematics and arrival time at near Earth.

After the launch of \textit{Solar TErrestrial RElation Observatory (STEREO)} in 2006 \citep{Kaiser2008}, it is possible to continuously image the CMEs and to determine the 3D locations of their features in the heliosphere and hence have direct evidence of CME interaction using images from Heliospheric Imager (HI) on Sun Earth Connection Coronal and Heliospheric Investigation (SECCHI) package \citep{Howard2008}. However, immediately after the launch of STEREO, during deep extended solar minimum, not many interacting CMEs were observed. As the current solar cycle progressed, interaction of CMEs now appears a fairly common phenomenon, in particular around solar maximum. The interacting CMEs of 2010 August 1 have been studied extensively by using primarily the STEREO white light imaging, near Earth in situ and, S-Waves radio observations \citep{Harrison2012, Liu2012, Mostl2012,Temmer2012,Martinez-Oliveros2012,Webb2013}. Also, \citet{Lugaz2012} have reported a clear deflection of 2010 May 23 - 24 CMEs after their interaction. As mentioned above, therefore, only a few studies on CME interaction have been reported. Several key questions regarding CME interaction need to be addressed, which are not well understood quantitatively,viz. (1) How the dynamics of CMEs change after interaction? What is the regime of interaction, i.e. elastic, inelastic or super-elastic? \citep{Lugaz2012,Shen2012,Mishra2014} (2) What are the consequences of the interaction of CME-shock structure? How the overtaking shock changes the plasma and the magnetic field properties into preceding magnetic cloud? \citep{Lugaz2005,Lugaz2012,Liu2012} (3) What are the favorable conditions for the CME cannibalism and the role of magnetic reconnection in it? \citep{Gopalswamy2001apj} (4) What is the possibility for production of reverse shock at CME-CME interaction site? \citep{Lugaz2005} (5) Whether these interacted structures produce different geomagnetic consequences than individual CMEs, on their arrival to magnetosphere? \citep{Farrugia2006}. (6) What are the favorable conditions for the CME's deflection and enhanced radio emission during CME-CME interaction? \citep{Lugaz2012,Martinez-Oliveros2012}. In light of aforementioned questions and only a few studies reported, it is clear that the prediction of arrival time of interacting CMEs and association of remote observations of such CMEs with in situ is challenging.

In the present study, we focus on the identification, evolution and propagation of two CMEs launched on 2010 November 9 and 10, as they travel from the corona to the inner heliosphere. This study, in which the launch time of the selected CMEs are separated by about 14 hr, gives an opportunity to understand the CME-CME interaction unambiguously for simpler scenario, contrary to complex events around 2010 August 1, wherein four CMEs were launched in quick succession. We find that CMEs of November 9 and 10 are directed towards the Earth and their initial characteristics indicate a probability of their interaction. In Section~\ref{IntCMEs}, observations and estimated kinematics of these CMEs are presented to investigate if they interacted during their propagation in the heliosphere. In Section~\ref{Momentum}, nature of collision, momentum and energy transfer during collision phase of CMEs are described using their estimated true mass and kinematics. In Section~\ref{Insitu}, in situ observations of remotely tracked CME features are described. Using the derived kinematics combined with Drag Based Model (DBM) of \citet{Vrsnak2013}, the arrival time and transit speed of tracked features of CMEs at L1 is predicted in Section~\ref{arrtime}. The geomagnetic response of interacted CMEs structures is described in Section~\ref{geomagnetic}. In Section~\ref{Results}, the results and discussions of the present study are emphasized and conclusions are given in Section~\ref{Conclu}.

\section{Observations}
\label{Obs}

\setlength{\emergencystretch}{1em} {For the present study of 2012 November 9-10, CME-CME interaction, we analyzed the white light images from SECCHI suite on-board NASA's twin STEREO (A \& B) mission.} SECCHI suite consists of an Extreme Ultraviolet Imager (EUVI), two coronagraphs (COR1 and COR2) and, two Heliospheric Imagers (HI1 and HI2). These instruments can image the heliosphere from near the Sun to the Earth and beyond.  The EUVI (Field-of-view (FOV): 1-1.7 $R_{\odot}$), COR1 (FOV: 1.5-4 $R_{\odot}$) and COR2 (FOV: 2.5-15 $R_{\odot}$) FOV are centered on the Sun. The HI1 and HI2 have their optical axis aligned in the ecliptic plane and are off-centered from the Sun at solar elongation of $14^{\circ}$ and $53.7^{\circ}$, respectively.  HI1 and HI2 have wide FOV of $20^{\circ}$ and $70^{\circ}$, respectively. With this identical imaging package (SECCHI) on twin spacecraft, a CME can be tracked from 0.4 to $88.7^{\circ}$ elongations. At the time of observations around November 9, the STEREO A \& B were $127^{\circ}$ westward and $123^{\circ}$ eastward from the Sun-Earth line at a distance of 0.96 AU and 1.08 AU from the Sun, respectively. The in situ properties were investigated through data collected by instruments on-board Wind spacecraft \citep{Lepping1995,Ogilvie1995}.

\subsection{Interaction of 2012 November 9-10 CMEs}
\label{IntCMEs}

In order to determine the true direction and speed of CMEs launched on November 9 and 10 in quick succession, 3D reconstruction technique is applied to CME features close to the Sun. We find that the CME of November 9 (hereinafter referred as, CME1) has a slow speed than fast speed November 10 CME (hereinafter referred as, CME2) and both move in a similar direction towards the Earth. These observations indicate the probability of their interaction in the heliosphere. Using SECCHI/HI observations, we attempt to investigate if these CMEs indeed interacted with each other in the heliosphere. We also attempt to identify various CME structures in the in situ observations taken at 1 AU.

\subsubsection{3D reconstruction in COR2 Field-of-view}
\label{3D}

A partial halo CME1 (angular width of $276^{\circ}$) was observed in SOHO/LASCO-C2 FOV at 15:12 UT on 2012 November 9 (\url{http://cdaw.gsfc.nasa.gov/CME_list/}). In the SECCHI/COR1-B and A, this CME was observed at 14:25 UT in the SW and SE quadrants, respectively. Another partial halo CME2 (angular width $\cong 210^{\circ}$) was observed in LASCO C2 FOV at 05:12 UT on November 10. This CME was detected by SECCHI/COR1-B and A in the SW and SE quadrants, respectively, at 05:05 UT on November 10, 2012. 

\begin{figure}
\begin{center}
\includegraphics[scale=0.60]{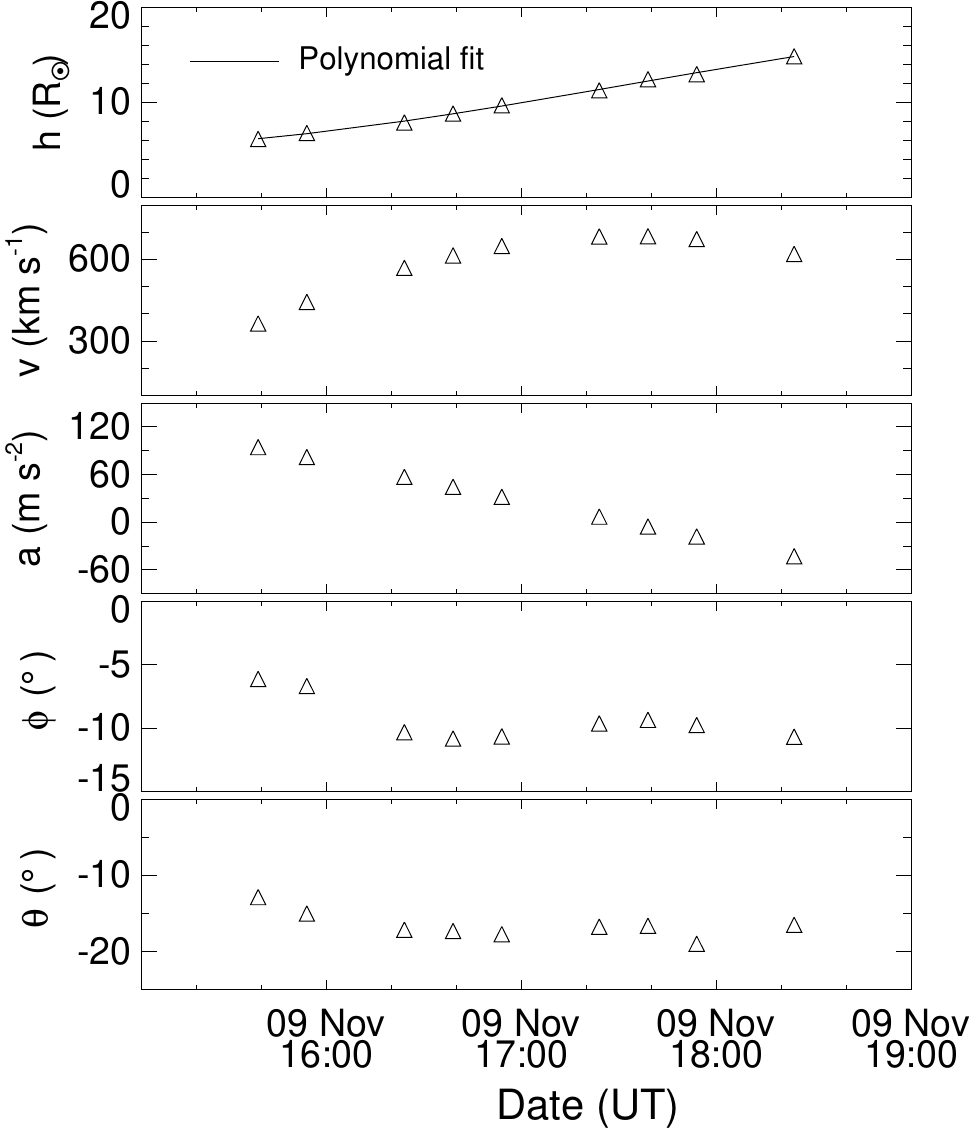}
\includegraphics[scale=0.60]{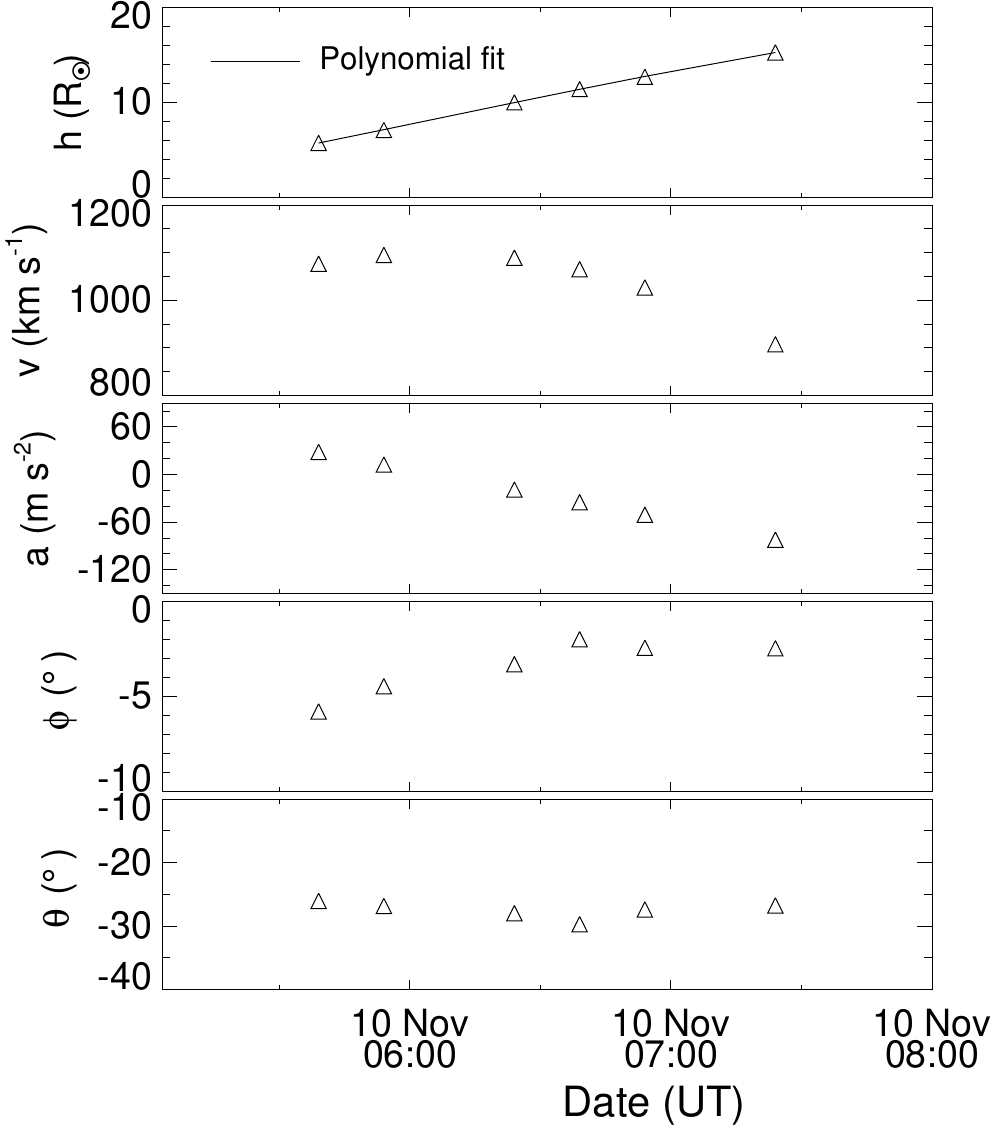}
\caption{\scriptsize{From top to bottom panels, true height, velocity, acceleration, longitude and latitude of selected feature along the leading edge as derived from Tie-Pointing method have been plotted as a function of time for CME1 (left) and CME2 (right)}}
\label{scc}
\end{center}
\end{figure}

\begin{figure}
\begin{center}
\includegraphics[scale=0.31]{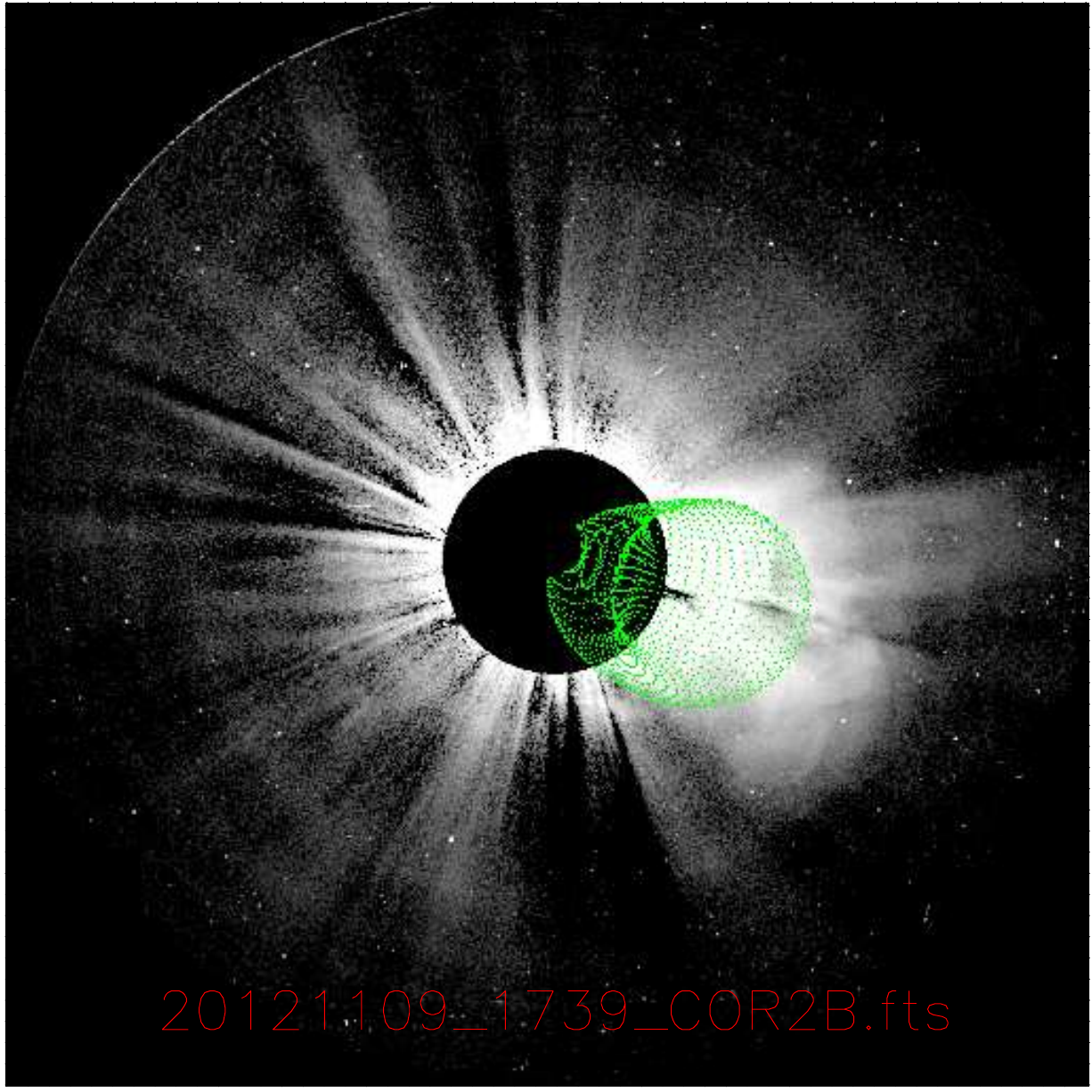}
\includegraphics[scale=0.31]{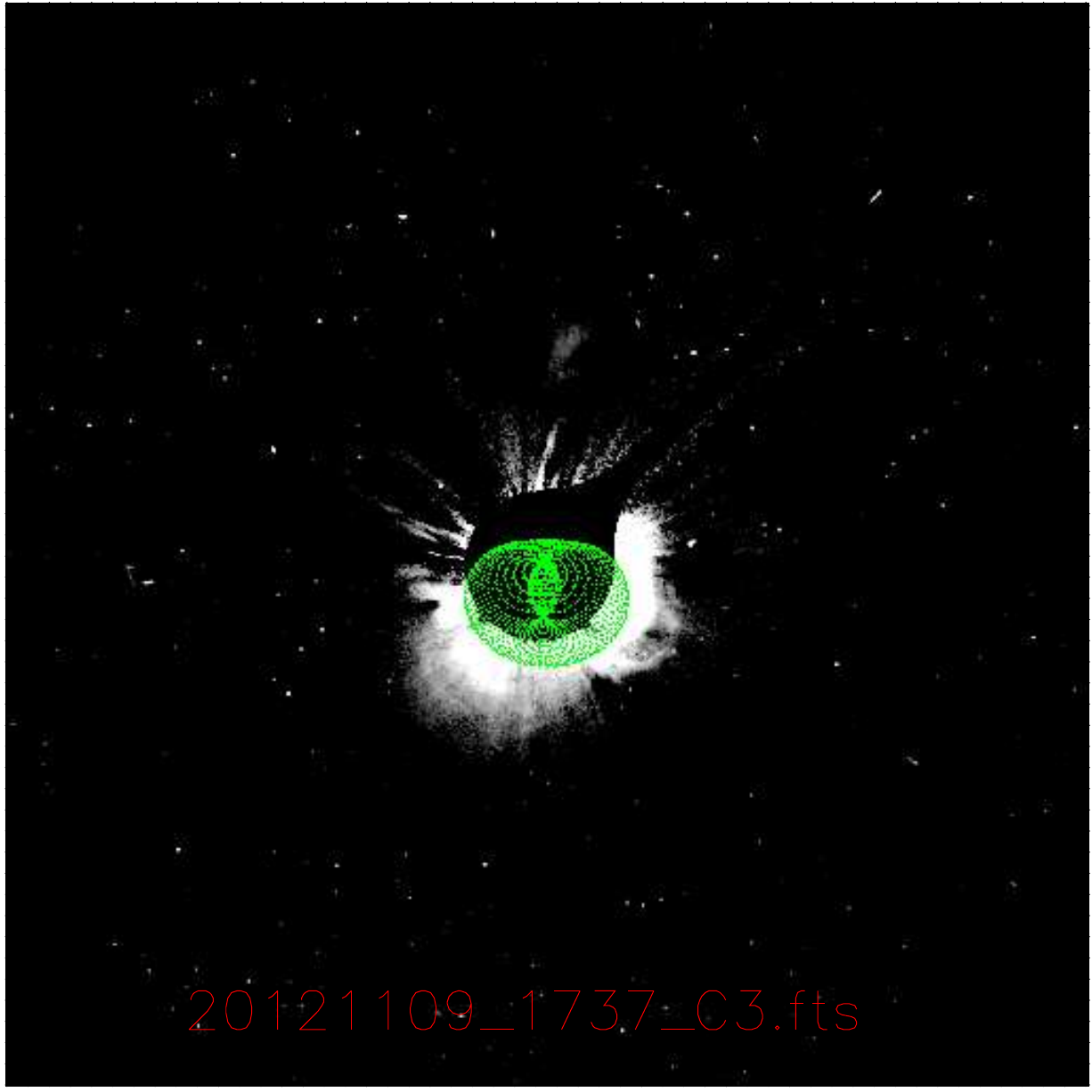}
\includegraphics[scale=0.31]{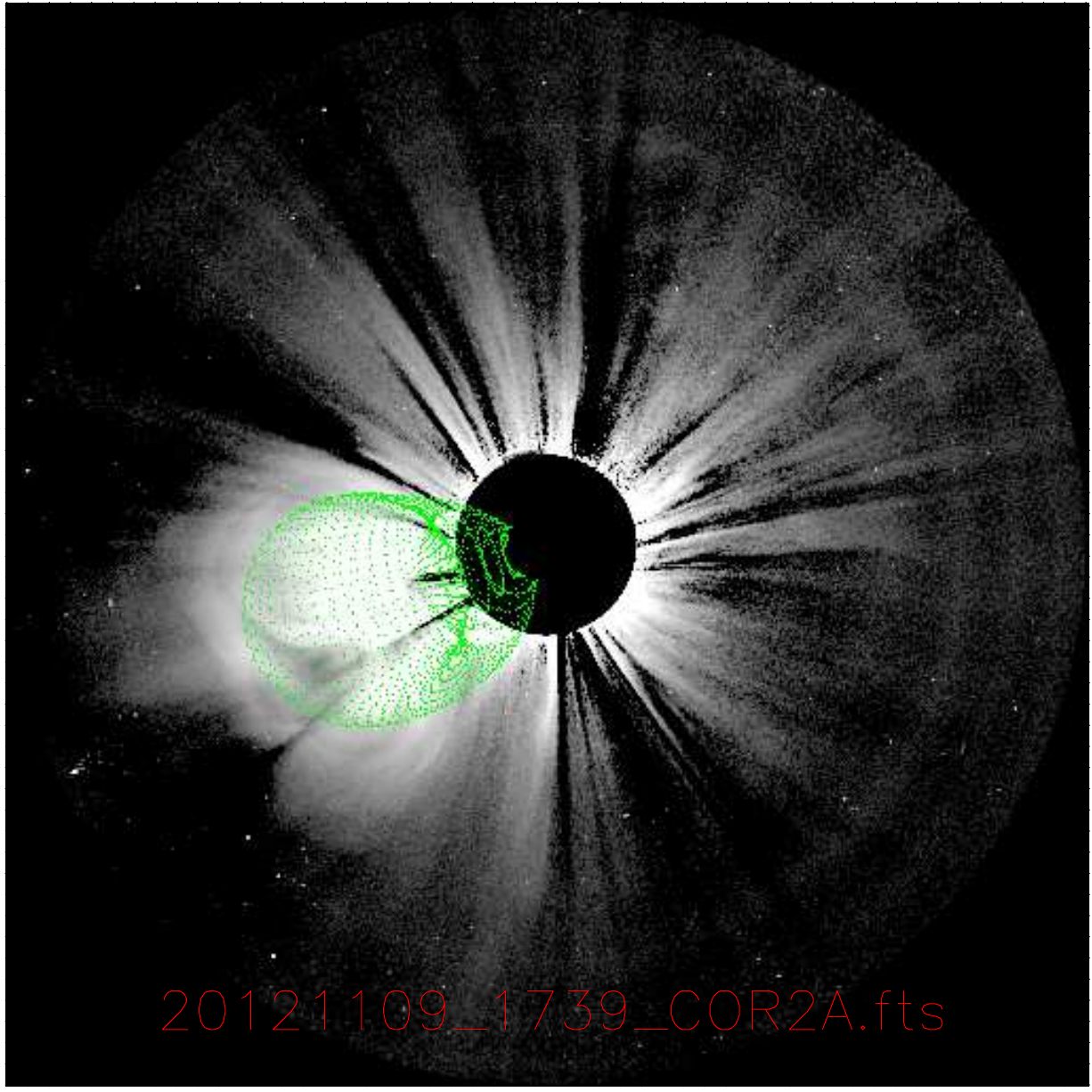}
\includegraphics[scale=0.31]{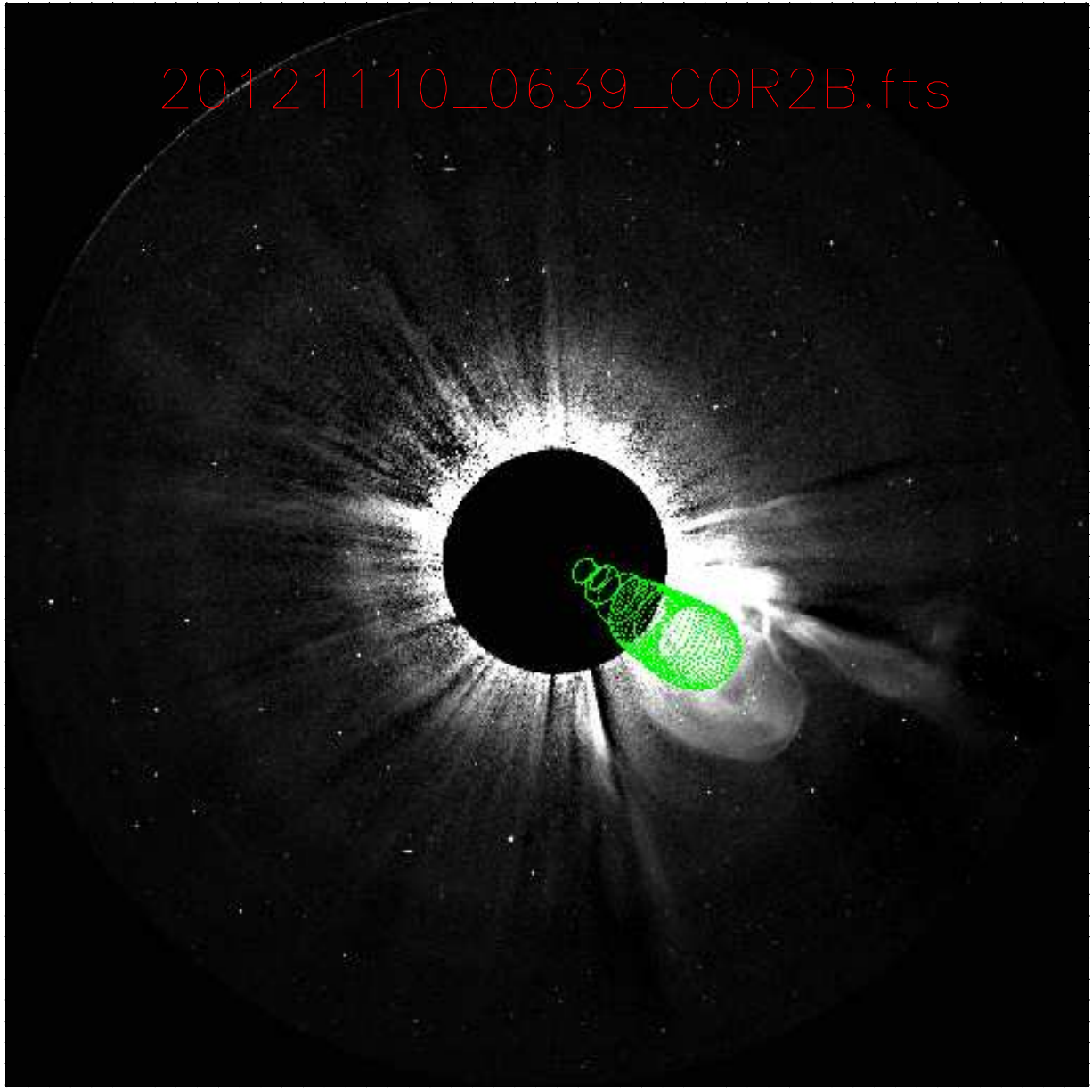}
\includegraphics[scale=0.31]{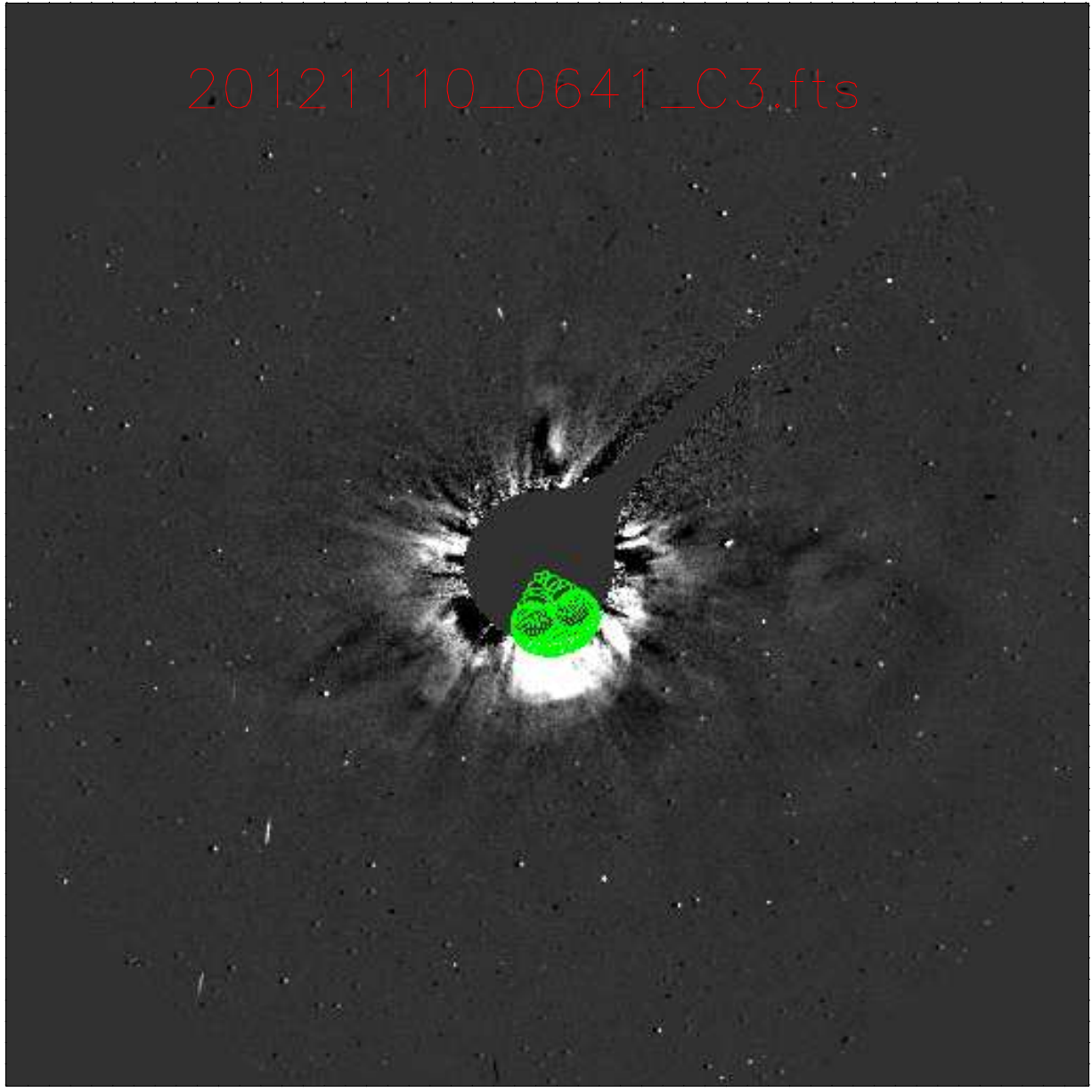}
\includegraphics[scale=0.31]{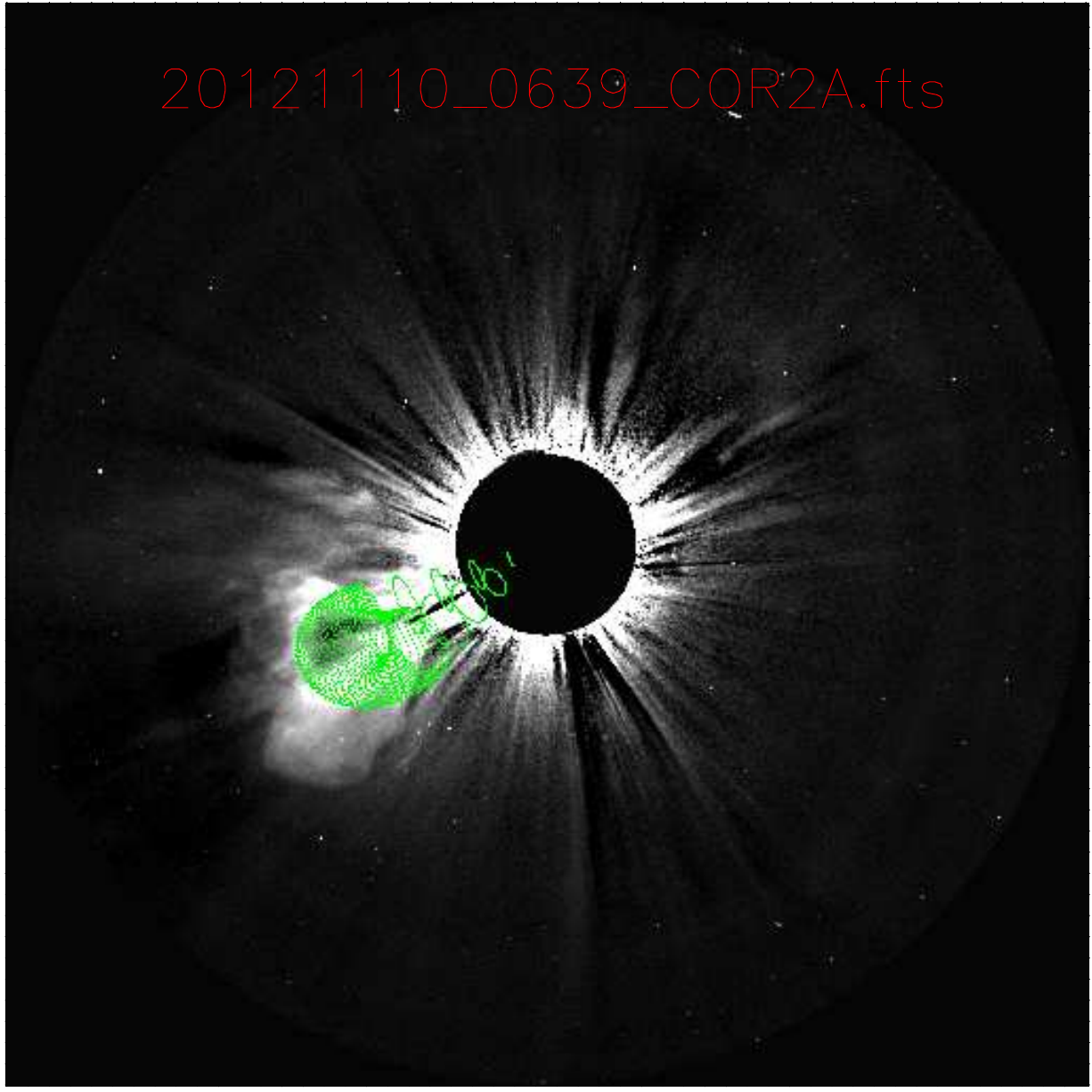}
\caption{\scriptsize{The top and bottom rows show the GCS model wireframe overlaid on the CME1 and CME2 images, respectively. The triplet of concurrent images around 17:39 UT on November 9 and around 06:39 UT on November 10 corresponding to CME1 and CME2, respectively, are from STEREO/COR2-B (left), SOHO/LASCO-C3 (middle) and STEREO/COR2-A (right).}}
\label{FM}
\end{center}
\end{figure}

Figure~\ref{scc} shows true kinematics of CME1 and CME2 from 3D reconstruction of selected features along the leading edges of corresponding CMEs using tie-pointing procedure (scc\_measure: \citealt{Thompson2009}) on SECCHI/COR2 data. This tie-pointing procedure uses the concept of epipolar geometry \citep{Inhester2006} and due to epipolar  constraints, the identification of same feature in two images (from STEREO-A and B) is always possible along the same epipolar line in both images. Before implementing this technique, the processing of SECCHI/COR2 images and creation of minimum intensity images and then its subtraction from the sequence of processed COR2 images was carried out following the approach described in \citet{Mierla2008,Srivastava2009}. As the separation angle of the STEREO spacecraft is large, it is better to use two independent methods to confirm the results of 3D reconstruction. Therefore, we have also used the visual fitting of both CMEs in the SECCHI/COR2 FOV using the Graduated Cylindrical Shell (GCS) model \citep{Thernisien2009,Thernisien2011} for the 3D reconstruction of these CMEs. This model represents the flux rope structure of CMEs with two shapes: the conical legs and the curved (tubular) fronts, also known as the ``hollow croissant''. We used contemporaneous image triplets of CME1 around 17:39 UT on November 9 from STEREO-A/COR2, STEREO-B/COR2 and SOHO/LASCO-C3. The best visual fit for CME1 is found in the direction of W02S14, with a half angle of 19$^{\circ}$, tilt angle of 9$^{\circ}$ around the axis of symmetry of the model (i.e. rotated 9$^{\circ}$ anticlockwise out of the ecliptic plane), and an aspect ratio of 0.52. At this time, CME1 was at a distance of about 9.6 $R_{\odot}$ from the Sun. Using the obtained fitted values of half angle and aspect ratio of CME1, its 3D face-on and edge-on angular width is estimated as 100$^\circ$ and 62$^\circ$, respectively. We also carried out the visual fitting using GCS model for CME2 around 06:39 UT on November 10 by exploiting the concurrent image triplets of STEREO-A/COR2, STEREO-B/COR2 and SOHO/LASCO-C3.  The best fit for CME2 is obtained in the direction of W06S25, at a tilt angle of $9^{\circ}$ around the axis of symmetry of the model, with a half angle of $12^{\circ}$, and an aspect ratio of 0.19. At this time CME2 was at a distance of 8.2 R$_{\odot}$ from the Sun. Using the GCS model fitted values, the 3D face-on and edge-on angular width of CME2 is estimated as 46$^\circ$ and 22$^\circ$, respectively. The GCS model fit for CME1 and CME2 is shown in the top and bottom panel of Figure~\ref{FM}, respectively. We find that the estimated latitude and longitude of both CMEs from GCS model and tie-pointing procedure are in agreement within few degrees. This shows that the results of both tie-pointing and GCS model are reliable and can be used for estimating kinematics of CMEs in the coronagraphic FOV.

From a study of 3D kinematics of the two CMEs using tie-pointing approach (scc\_measure: \citealt{Thompson2009}), we found that CME1 is slow with a true speed of 620 km s$^{-1}$ around 15 R$_{\odot}$ while CME2 is faster with a speed of 910 km s$^{-1}$ at a distance of approximately 15 R$_{\odot}$ (Figure~\ref{scc}). Latitude and longitude of reconstructed features for both CMEs suggest that these were Earth-directed and could possibly interact in the interplanetary (IP) medium. If the speed of CMEs is assumed to be constant beyond the outer edge of COR2 FOV, then these CMEs should have collided at approximately 130 R$_{\odot}$ on November 11 around 02:30 UT. However, previous studies have shown that the speed of a CME can significantly change after the coronagraphic FOV \citep{Lindsay1999,Gopalswamy2000,Cargill2004,Manoharan2006,Vrsnak2010}. Therefore, further tracking of CME features in the heliosphere is required to determine the exact location and time of interaction of these CMEs.

\subsubsection{Tracking of CMEs in HI Field-of-view}
\label{HI}

\begin{figure}
\begin{center}
\includegraphics[scale=0.10]{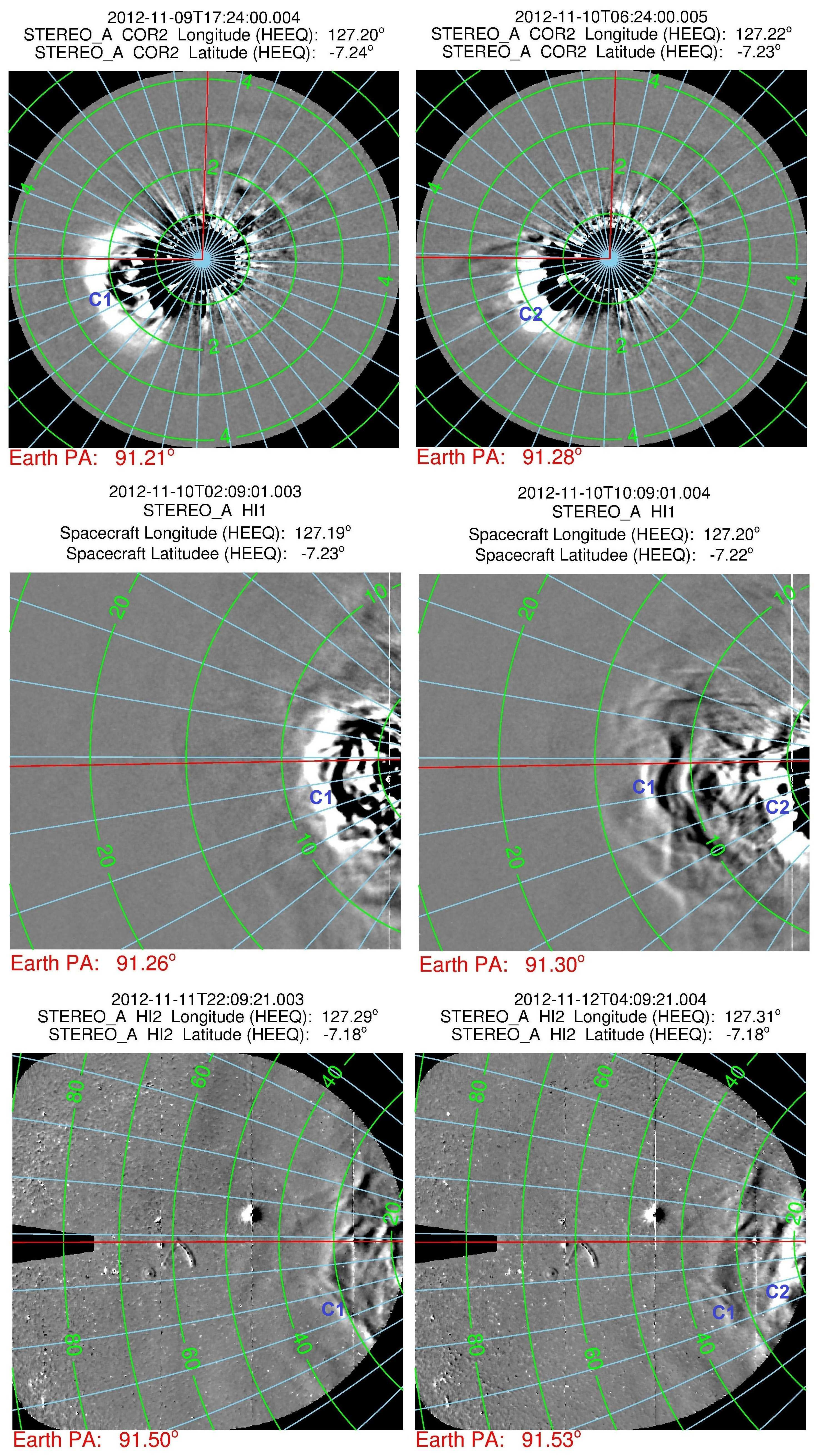}
\caption{\scriptsize{Evolution of CMEs in running difference images of COR2, HI1 and HI2 FOV is shown in upper, middle and lower panels respectively. Left and right panels show observations from STEREO/SECCHI Ahead spacecraft at two different times. Contours of elongation (green) and position angle (blue) is overlaid on the images. In each image, the postion angle is overlaid in interval of $10^{\circ}$ and the horizontal red line is along the ecliptic at the position angle of Earth. In upper panel (both left and right) the vertical red line marks the $0^{\circ}$ position angle. In each panels, the C1 and C2 corresponds to CME1 and CME2.}}
\label{evolution}
\end{center}
\end{figure}

\begin{figure}
\begin{center}
\includegraphics[scale=0.5]{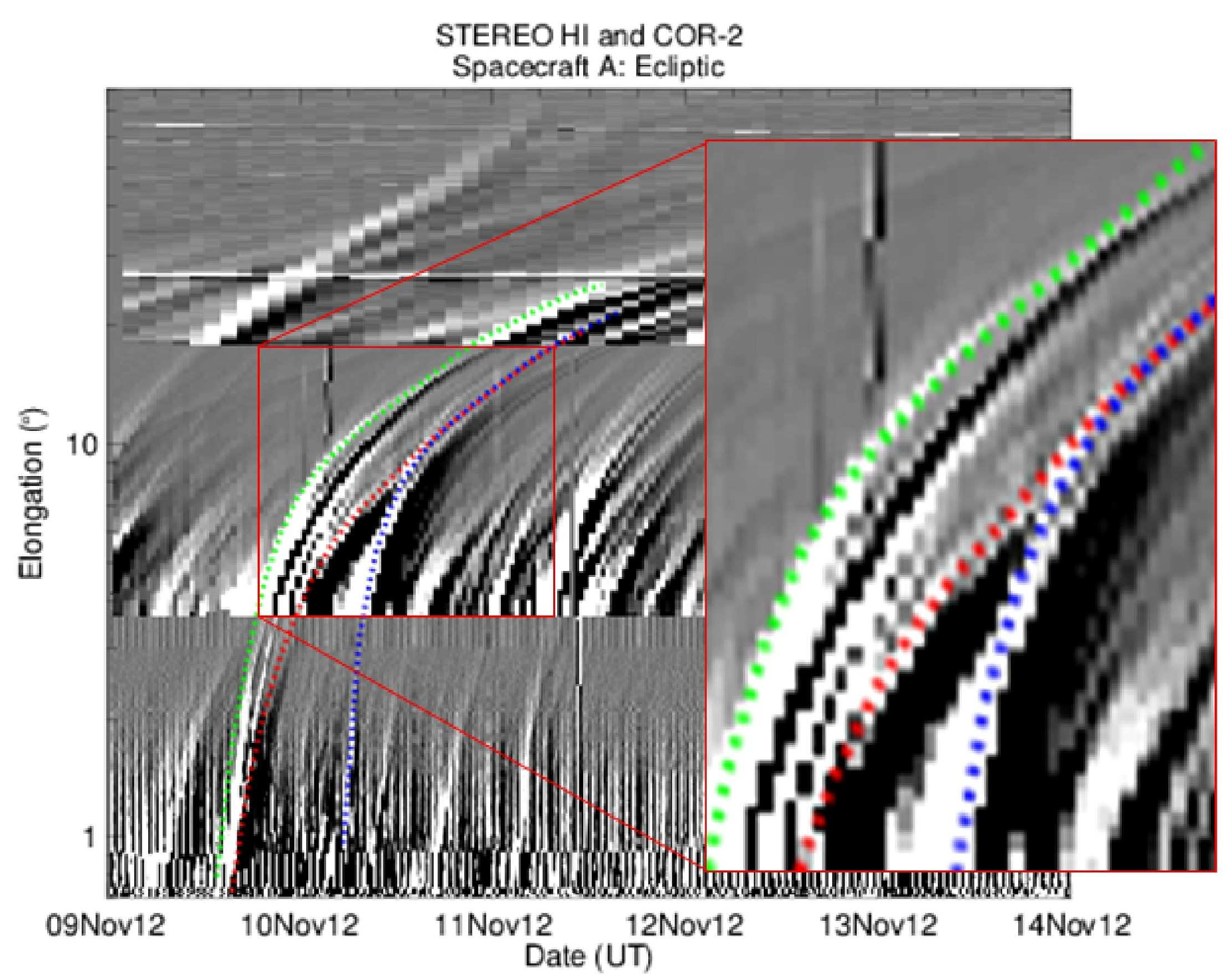}
\caption{\scriptsize{Time-elongation map (J-map) using the COR2 and HI observations of STEREO/SECCHI spacecraft during interval of 2012 November 9-14 is shown. The features corresponding to CME1 leading edge (LE), CME1 trailing edge (TE) and CME2 leading edge are (LE) tracked and over plotted on the J-map with green, red and blue, respectively.} The red rectangle (rightmost) is an enlarged plot of the red rectangle (on the left) which clearly shows that the red and blue tracks meet in HI1 FOV.}
\label{J-maps}
\end{center}
\end{figure}

\begin{figure}
\begin{center}
\includegraphics[height=6cm, width=3.9cm]{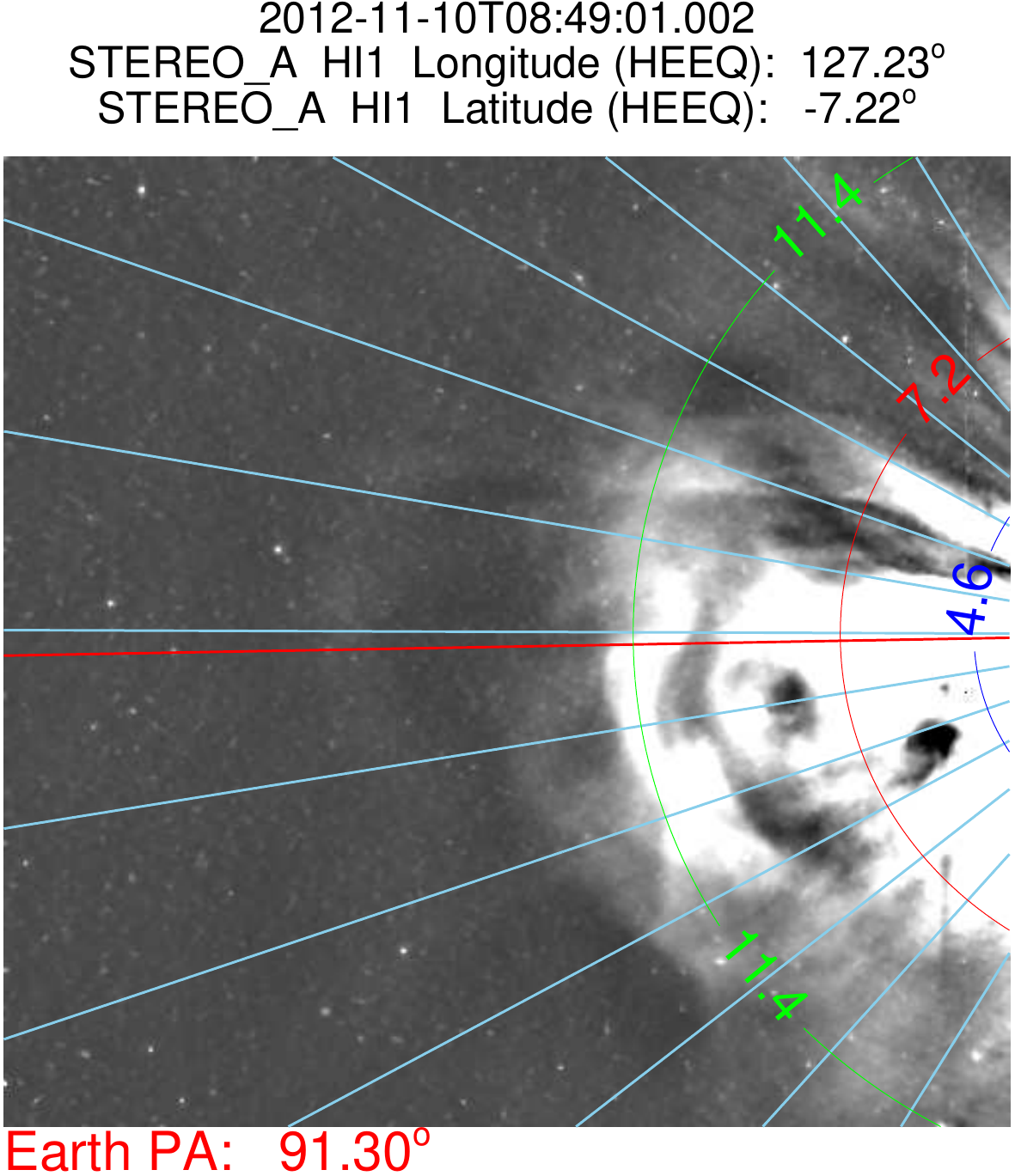}
\includegraphics[height=6cm, width=3.9cm]{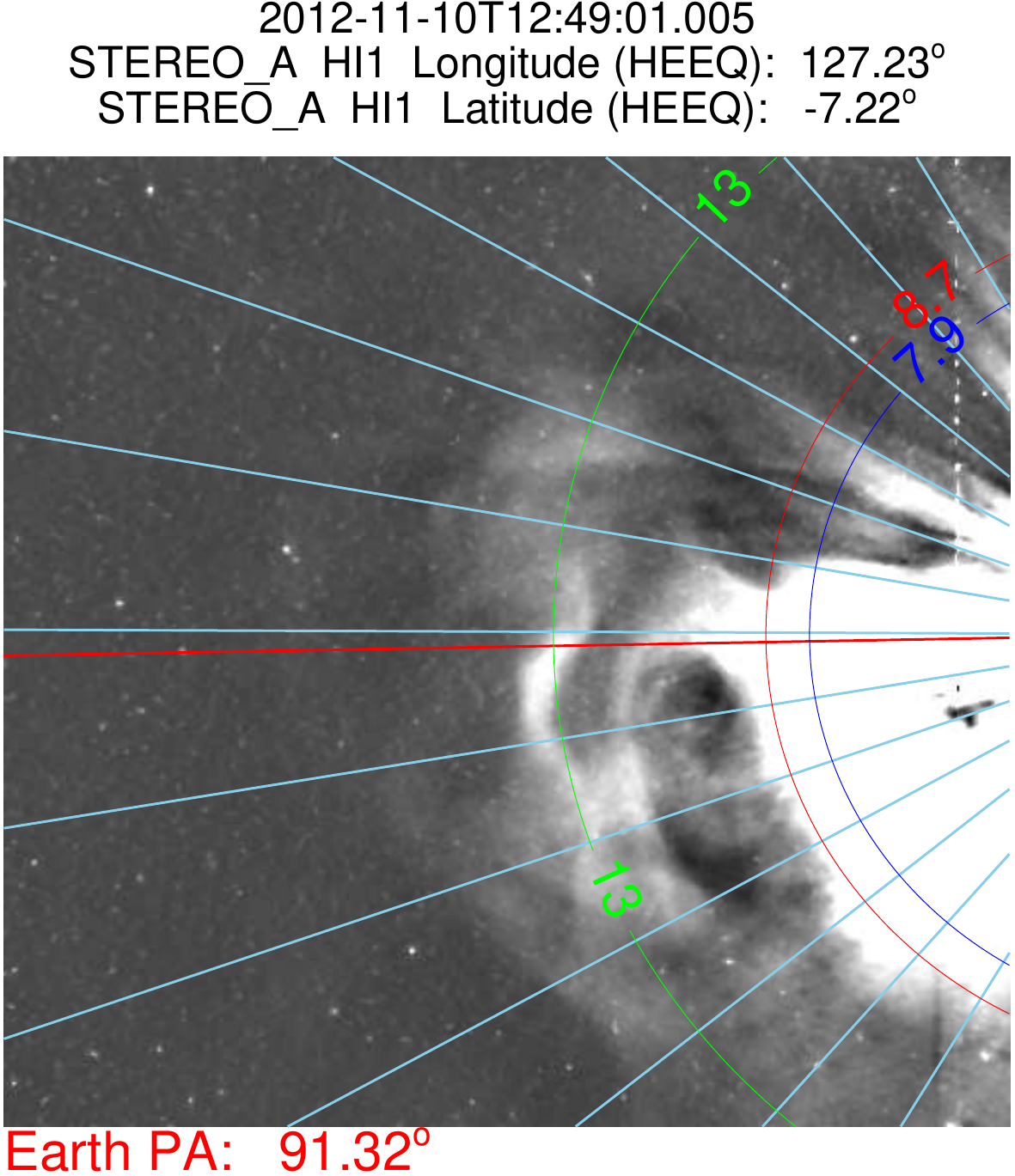}
\includegraphics[height=6cm, width=3.9cm]{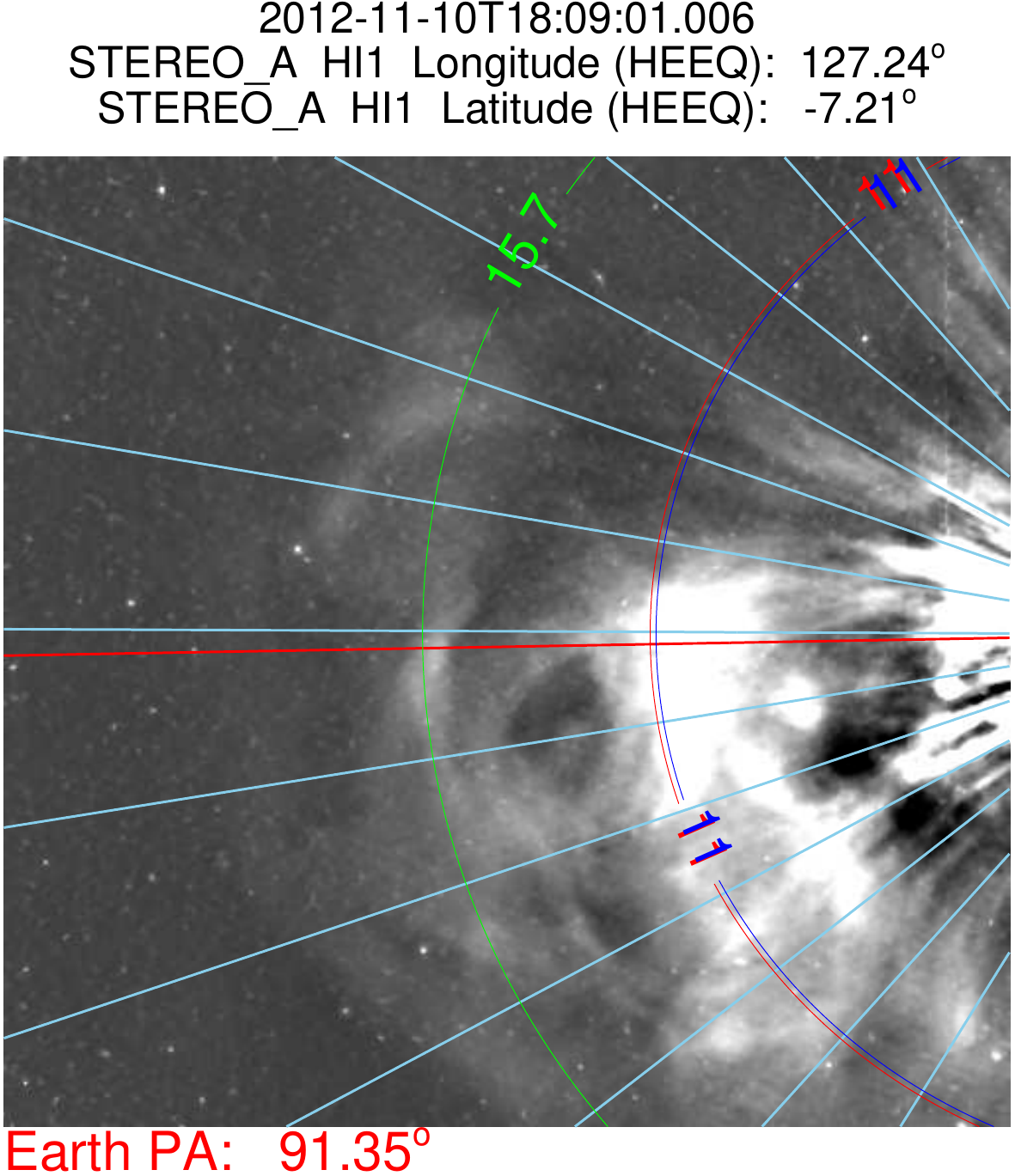}
\caption{\scriptsize{The contours of derived elongation of tracked CME1 LE (green), CME1 TE (red) and CME2 LE (blue) features from the J-map is overplotted on the base difference HI1-A images. In each image, the position angle (sky blue) is overlaid in interval of $10^{\circ}$ and the horizontal red line is along the ecliptic at the position angle of Earth.}}
\label{basediff}
\end{center}
\end{figure}

Evolution of CME1 and CME2 in the running difference images of COR2, HI1 and HI2 FOV is shown in Figure~\ref{evolution}. To track and estimate the arrival times of the CMEs in the heliosphere using HI1 \& 2 images, we constructed J-maps based on the method developed by \citet{Sheeley1999,Davies2009} and derived the variation of elongation of selected features with time. The details about extraction of strip of constant position angle, selection of bin size and adopted procedure for construction of J-maps and derivation of elongation angle from it has been described in section 3.1.1 of \citet{Mishra2013}. The positively inclined bright features in the J-maps (Figure~\ref{J-maps}) correspond to enhanced density structure of the CMEs. We tracked the leading and trailing edge (marked in green and red) of bright features corresponding to slow CME1 and leading edge (blue) of bright feature corresponding to the fast CME2. Prior to interpreting the tracked features of CME1 in J-map as its leading and trailing edge, respectively, the derived elongations corresponding to all the three tracked features were overplotted on the base difference HI1 images. On careful inspection of the sequence of these images, we noticed that the two tracked features of CME1 correspond to first density enhancement at the front and second density enhancement at the rear edge of CME1, respectively. The base difference HI1-A images with overplotted contours of elongation corresponding to tracked features of CME1 and CME2, are shown in Figure~\ref{basediff}. It is to be noted that the term `trailing edge' of CME1 does not correspond to the rear most portion of CME1 but corresponds to the part behind the front and is not an image artifact in J-map constructed from running difference images.

Since, the tracked features of CME2 were not observed well in STEREO B ecliptic J-map, we could not implement the stereoscopic reconstruction technique to estimate the CME kinematics. Instead, we used Harmonic Mean (HM) approximation which  requires single spacecraft observation to derive the elongation variations with time for the tracked features which could then be converted into heliocentric distances from the Sun \citep{Lugaz2009}. In the HM approximation, geometry of a CME is considered as a self-similar expanding sphere attached to the Sun and its centre moves along a fixed longitude. The choice of the HM method is based on our earlier study on comparison of various reconstruction techniques involving both stereoscopic and single spacecraft observations \citep{Mishra2014.apj}. It was found that amongst the single spacecraft techniques, the HM method is the most suitable for arrival time prediction. Implementing the 3D reconstruction technique in Section~\ref{3D} (Figure~\ref{scc}), we have shown that the estimated longitude of CME1 and CME2 are approximately 10$^{\circ}$ and 2$^{\circ}$ east of the Sun-Earth line. We assume that beyond the COR2 FOV, CMEs continue to propagate along the same direction i.e. we ignore any heliospheric longitudinal deflection of CMEs. Therefore, the aforesaid estimated values of longitude were used in the HM approximation to convert the derived elongation from J-maps to radial distance from the Sun.

\begin{figure}
\begin{center}
\includegraphics[scale=0.75]{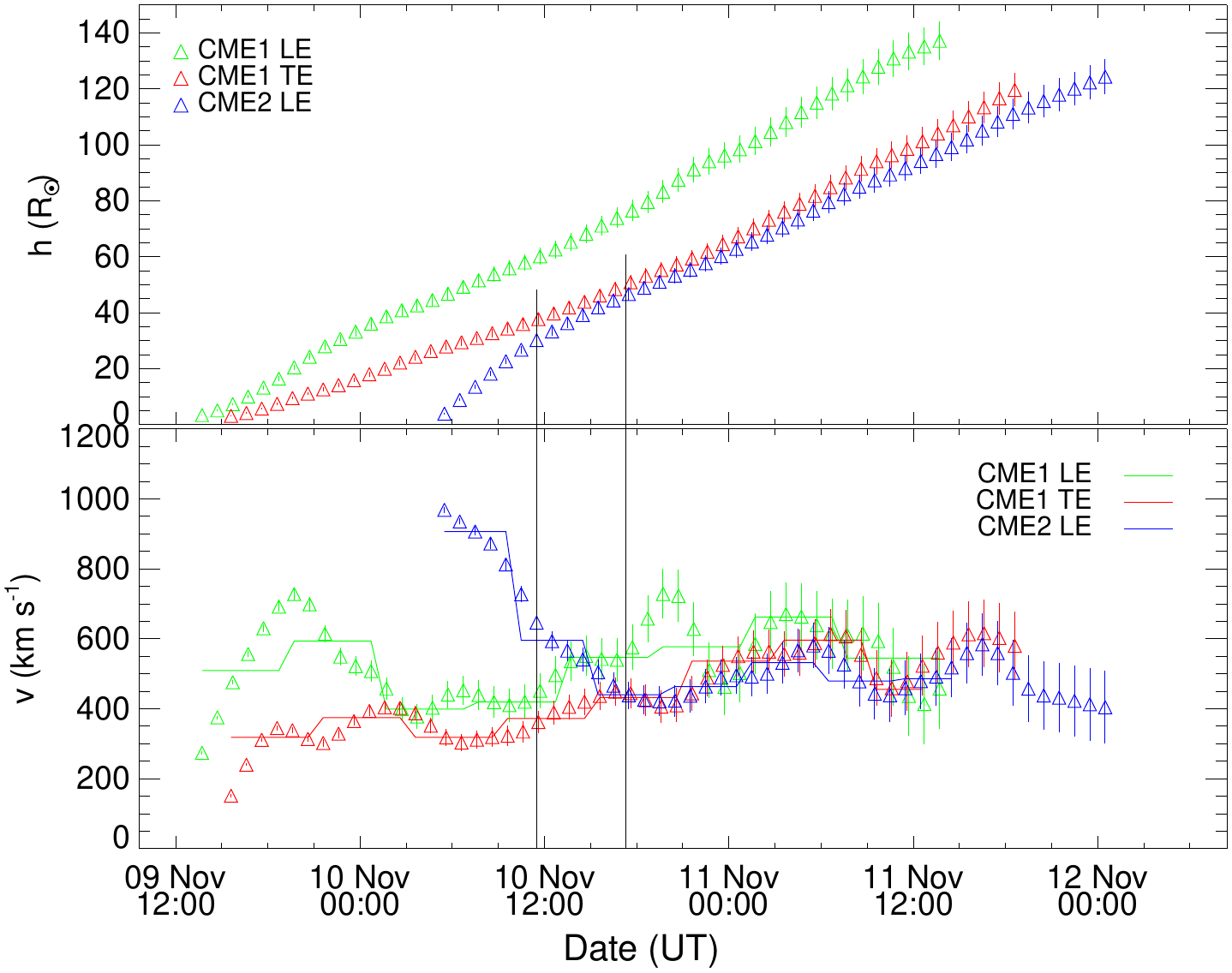}
\caption{\scriptsize{In the top and bottom panels, distance and speed of the tracked features CME1 LE, CME1 TE and CME2 LE as marked in Figure~\ref{J-maps} with green, red and blue, respectively, are shown. Speeds of these features are calculated from the differentiation of adjacent distance points using the three point Lagrange interpolation (shown with $\Delta$ symbol in the bottom panel). Speed of these features are estimated by differentiating the first order polynomial fit to estimated distance points for an interval of approximately 5 hr (shown in solid line in the bottom panel). In the top and bottom panels, vertical lines show the error bars. We have assumed a fractional error of 5\% in estimated distance which is used to determine the uncertainties in the speeds. From the left, the two vertical lines (black) mark the start and end of the collision phase.}}
\label{kinematics}
\end{center}
\end{figure}

Figure~\ref{kinematics} shows the distance and speed estimated for different tracked features of the two CMEs. The estimated speed is derived from the adjacent distances using a numerical differentiation with three point Lagrange interpolation and therefore has systematic variations due to slight errors in the estimated distance. It is clear that the leading edge (LE) of CME1 has higher speed ($\approx$ 500 km s$^{-1}$) than its trailing edge (TE) speed ($\approx$ 350 km s$^{-1}$), averaged over few data points at the entrance of HI1 FOV. Also, the LE and TE of CME1 have lower speeds than LE of CME2. LE of CME2 shows a large radial speed of approximately 1100 km s$^{-1}$ (ecliptic speed = 950 km s$^{-1}$) in the COR2 FOV i.e.(2.5-15 R$_{\odot}$) (Figure~\ref{scc}). Beyond 10 R$_{\odot}$ distance, the LE of CME2 continuously decelerates for $\approx$ 10 hours up to 46 $R_{\odot}$ where its speed reduces to 430 km s$^{-1}$. The extreme fast deceleration of LE of the CME2, starting from COR2 FOV, seems to be due to possible interaction with the preceding CME1. It is likely that the CME1 had large spatial scale due to which the trailing plasma and magnetic fields from CME1 created sufficiently dense ambient medium acting as a huge drag force for CME2 which resulted in its observed deceleration. The extreme fast deceleration of CME2 LE can also be due to closed magnetic structure of CME1 which may act like a magnetic obstacle for the CME2 \citep{Temmer2012}.

From estimated kinematics of tracked features in heliosphere (top and bottom panel of Figure~\ref{kinematics}), it is clear that around November 10 at 11:30 UT speed of CME1 TE started to increase from 365 km s$^{-1}$ with simultaneous decrease of CME2 LE speed from 625 km s$^{-1}$. Such an observation of acceleration of CME1 and deceleration of CME2 provides an evidence for the commencement of collision \citep{Temmer2012,Lugaz2012,Shen2012,Maricic2014,Temmer2014,Mishra2014}. We define collision phase, as the interval during which the two CMEs come in close contact with each other and show opposite trend of acceleration relative to one another until they attain an approximately equal speed or their trend of acceleration is reversed. Carefully  observing the variations of speed of tracked CME1 TE and CME2 LE, the start and end boundary of collision phase is drawn as vertical lines in the bottom panel of Figure~\ref{kinematics}. We noticed that at the end of the collision phase around November 10 at 17:15 UT, speed of CME1 TE is $\approx$ 450 km s$^{-1}$ and speed of CME2 LE is 430 km s$^{-1}$. In the beginning of the marked collision phase, CME1 TE is at a distance of 37 R$_\odot$ and CME2 is at 30 R$_\odot$. At the end of the marked collision phase, the CME1 TE  and CME2 LE are at a distance of 50 R$_\odot$ and 46 R$_\odot$, respectively. Here, we must highlight that throughout the collision phase, the estimated heliocentric distance of CME1 TE and CME2 LE is never found to be equal. This is because of the large scale structure of CMEs and we suggest that the tracked TE of CME1 and LE of CME2 using J-maps, are not strictly the rearmost trail of CME1 and outermost front of CME2, respectively. Therefore, we had defined `collision' when tracked features of CMEs (i.e. time-elongation tracks) come in close contact (separation of $\approx$ 5 R$_\odot$) with one another. Such small separation between observed collided features is possible if forward shock driven by CME2 interacts with the rear most portion of the CME1 as also suggested by \citet{Temmer2012,Maricic2014}.

\subsection{Momentum, Energy Exchange and Nature of Collision}
\label{Momentum}
To understand the momentum exchange during collision of CMEs, we require their true masses. Since, appearance of a CME is due to Thomson scattered photospheric light from the electrons in the CME \citep{Minnaert1930,Billings1966,Howard2009}, therefore, the recorded scattered intensity can be converted into the number of electrons and hence mass of a CME can be estimated, if composition of CME is known. But an observer from different vantage points receives different amount of scattered light by the electrons, therefore, the true location of electrons of CME, i.e. propagation direction of CME must be known to estimate the true mass of CME. For a long time, mass of a CME was calculated assuming the CME location in the observer's plane of sky \citep{Munro1979,Poland1981,Vourlidas2000}. Although, the propagation direction of CME is calculated in our study using tie-pointing \citep{Thompson2009} and Forward modeling \citep{Thernisien2009} methods and described in Section~\ref{3D}, but to avoid any bias of the reconstruction methods, we use the method of \citet{Colaninno2009} which is based on Thomson scattering theory, to estimate the true propagation direction and true mass of both the CMEs. We used the simultaneous image pair of SECCHI/COR2 to estimate the true mass of CMEs. First, we estimated the mass of CME i.e. M$_{A}$ and M$_{B}$, from STEREO-A and STEREO-B locations. Then we took the ratio of equations (7) and (8) of \citet{Colaninno2009} and obtained a new expression as
$M_{A}/M_{B} = B_{e}(\theta_{A})/B_{e}(\theta_{A} + \Delta)$,
where $\theta$$_{A}$ is the angle of propagation direction of CME measured from the plane-of-sky (POS) of \textit{STEREO-A}, B$_{e}$($\theta$$_{A}$) is the brightness of a single electron at an angular distance of $\theta$$_{A}$ from the POS and $\Delta$ is the separation angle between both \textit{STEREO-A} and \textit{STEREO-B} from the Sun-Earth line. Using the measured M$_{A}$ and M$_{B}$ in aforementioned expression, the value of $\theta_{A}$ can be estimated. However, we noticed that multiple values of $\theta_{A}$ correspond to same ratio of M$_{A}$ and M$_{B}$ but the correct value of $\theta_{A}$ among them can be selected by visual inspection of CME in COR2 images. We obtained the true mass of CME using the derived value of $\theta_{A}$ in equation (4) of \citet{Colaninno2009}. The estimated mass of CME1 at 18:39 UT on November 9 at $\approx$ 15 R$_{\odot}$, from STEREO-A and STEREO-B vantage point is 4.60 $\times$ 10$^{12}$ kg and 2.81 $\times$ 10$^{12}$ kg, respectively. For CME2, the estimated mass at 07:24 UT on November 10 at $\approx$ 15 R$_{\odot}$, from STEREO-A and STEREO-B locations is 2.25 $\times$ 10$^{12}$ kg and 1.31 $\times$ 10$^{12}$ kg, respectively. The estimated propagation direction of CME1 and CME2 is 19$^\circ$ and 21$^\circ$ West from the Sun-Earth line, respectively. We estimated the true mass of CME1 and CME2 as M$_{1}$ = 4.66 $\times$ 10$^{12}$ kg and M$_{2}$ = 2.27 $\times$ 10$^{12}$ kg, respectively.

We assume that after crossing the COR2 FOV and during collision of CME1 and CME2, their estimated true masses (M$_{1}$ and M$_{2}$) remain constant as estimated earlier. From the Figure~\ref{kinematics}, the observed velocity of CME1 and CME2 before the collision is estimated as (u$_{1}$,u$_{2}$) = (365,625) km s$^{-1}$ and observed velocity of CME1 and CME2 after the collision is (v$_{1}$,v$_{2}$) = (450,430) km s$^{-1}$. To understand the nature of collision of large scale magnetized plasmoids, we attempt to estimate the coefficient of restitution ($e$) of colliding CME1 and CME2. The coefficient of restitution measures the bounciness (efficiency to rebound) of a pair of objects in collision and is defined as ratio of their relative velocity of separation to relative velocity of approach. Hence, for $e$ $<$ 1, $e$ = 1, and $e$ $>$ 1 the collision is termed as inelastic, elastic and super-elastic and consequently the kinetic energy of the system after the collision is found to decrease, equal and increase than before the collision, respectively. We restrict ourselves not to estimate the value of $e$ directly by using the pre and post collision velocity of CMEs. This is because the velocities estimated from reconstruction method have some errors and do not guarantee conservation of momentum, a necessary condition of collision, therefore, can lead to erroneous estimation of $e$ value. In our approach, considering the condition of conservation of total momentum for the collision, the velocity of CME1 and CME2 after the collision can be estimated theoretically (v$_{1th}$,v$_{2th}$), provided that velocity of CME1 and CME2 before the collision and the coefficient of restitution ($e$) value are known.

Using the velocities (u$_{1}$,u$_{2}$) and masses (M$_{1}$,M$_{2}$), we iterate for a range of $e$ values and estimate the (v$_{1th}$,v$_{2th}$) each time. Using this approach, the most suitable value of $e$, using which the theoretically estimated (v$_{1th}$,v$_{2th}$) is found to be closest to the observed (v$_{1}$,v$_{2}$) i.e. for which the variance,\\
 $\sigma = \sqrt{(v_{1th} - v_{1})^{2} + (v_{2th} - v_{2})^{2}}$, is minimum, can be obtained. The details of the above approach and the equations used, are explained in \citet{Mishra2014}. Using the aforementioned approach, the value of $e$ = 0.1 is found with (v$_{1th}$,v$_{2th}$) = (458,432) km s$^{-1}$ and $\sigma$ = 9. With this theoretically estimated $e$ value, total kinetic energy of CMEs is found to decrease by 6.7\% of its value before the collision. Using the estimated velocities of the CMEs before and after the collision (Figure~\ref{kinematics}), the value for $e$ is calculated equal to 0.08, which is approximately same as obtained from iterations described above. Therefore, our analysis suggests that the observed collision between the CMEs are close to perfectly inelastic in nature. The kinetic energy of CME1 and CME2 before the collision was 3.1 $\times$ 10$^{23}$ joules and 4.4 $\times$ 10$^{23}$ joules, respectively. After the collision, based on the observed speeds, we found that the kinetic energy of CME1 increased by 51\% to its value before the collision while the kinetic energy of CME2 decreased by 54.5\% to its value before the collision. We also noticed that after the collision, momentum of CME1 increased by 23\% and momentum of CME2 decreased by 31\% to their values before the collision. Such calculations support the claim that significant exchange of kinetic energy and momentum take place during CME-CME collision \citep{Temmer2012,Lugaz2012,Maricic2014,Shen2012}.

\begin{figure}
\begin{center}
\includegraphics[scale=0.75]{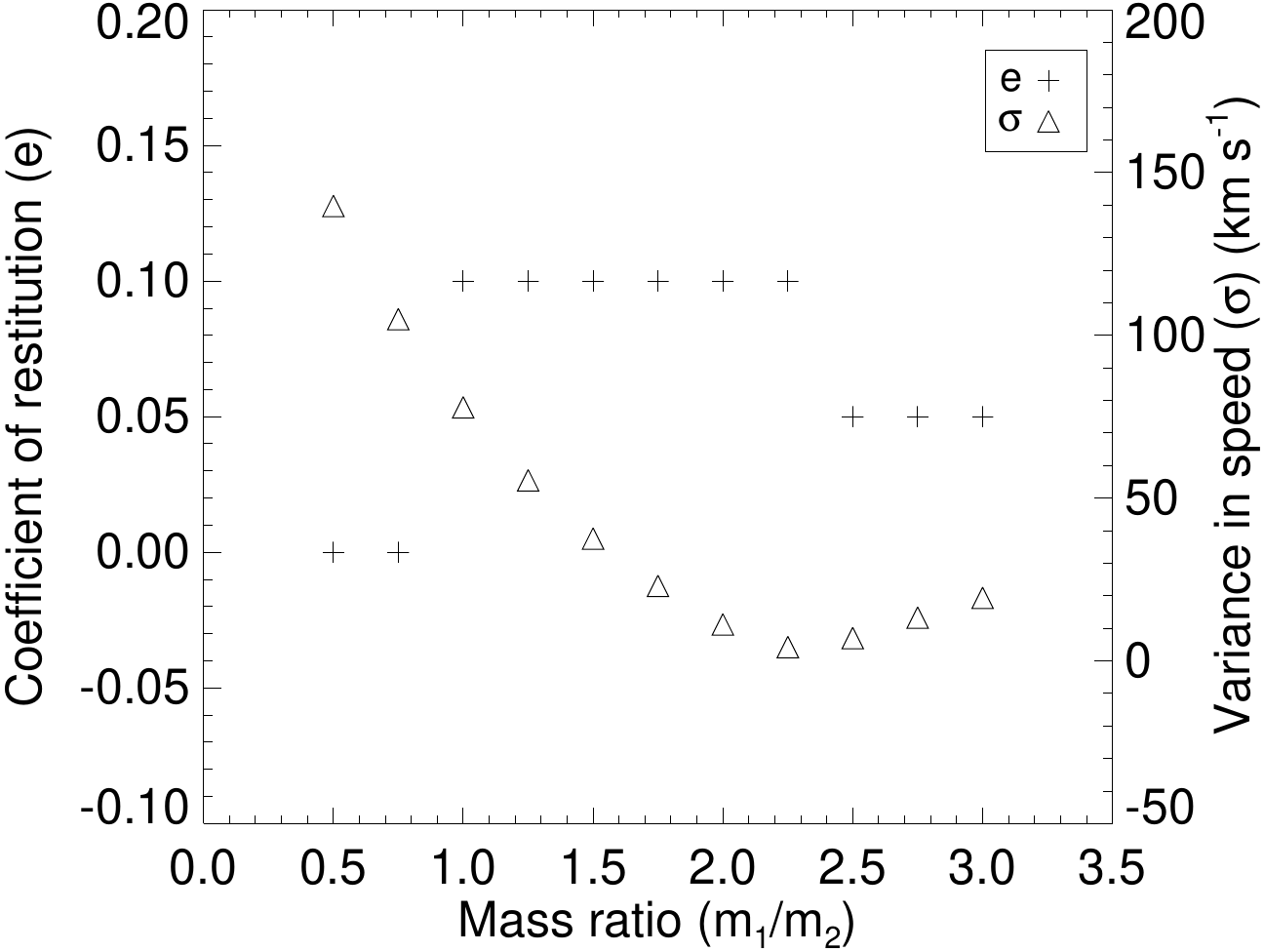}
\caption{\scriptsize{The best suited coefficient of restitution ($e$) corresponding to different mass ratios of CME1 and CME2 are shown with $+$ symbol and corresponding variance ($\sigma$) in velocity is shown with $\Delta$ symbol.}}
\label{mass}
\end{center}
\end{figure}

Considering the uncertainties of $\pm$ 100 km s$^{-1}$ in the speed, we repeated the above analysis. Taking the (v$_{1}$,v$_{2}$) = (550,530), we found $e$ = 0, i.e. perfectly inelastic. But in this case, (v$_{1th}$,v$_{2th}$) is estimated as (450,450) km s$^{-1}$ which highlight that theoretically estimated velocity is not close to the observed velocity (v$_{1}$,v$_{2}$). If we consider the (v$_{1}$,v$_{2}$) = (350,330), then it implies that the velocities of both CMEs after the collision are less than their values before the collision. Therefore, such velocities also cannot be considered for our case (as defined in Section~\ref{HI}). As, there are many sources of errors in the estimation of true mass of CMEs, therefore, the effect of uncertainty in mass must also be examined before confirming the nature of observed collision. In our case, the mass ratio (M$_{1}$/M$_{2}$) is equal to 2.05, varying this ratio  between 0.5 to 3.0 in a step of 0.25, and repeating the iterative procedure for estimating the most suitable value of $e$ corresponding to each mass ratio, we obtain $e$ close to 0. Figure~\ref{mass} shows the variation of $e$ and $\sigma$ with varying mass ratio. It is clear that despite taking large uncertainties in the mass of CME1 and CME2, the nature of observed collision remains close to perfectly inelastic for the case of interacting CMEs of 2012 November 9-10.

\section{In Situ observations, Arrival time and Geomagnetic Response of 2012 November 9-10 CMEs}

\subsection{In Situ Identification of Tracked CME Features}
\label{Insitu}
\begin{figure}
\begin{center}
\includegraphics[scale=0.65]{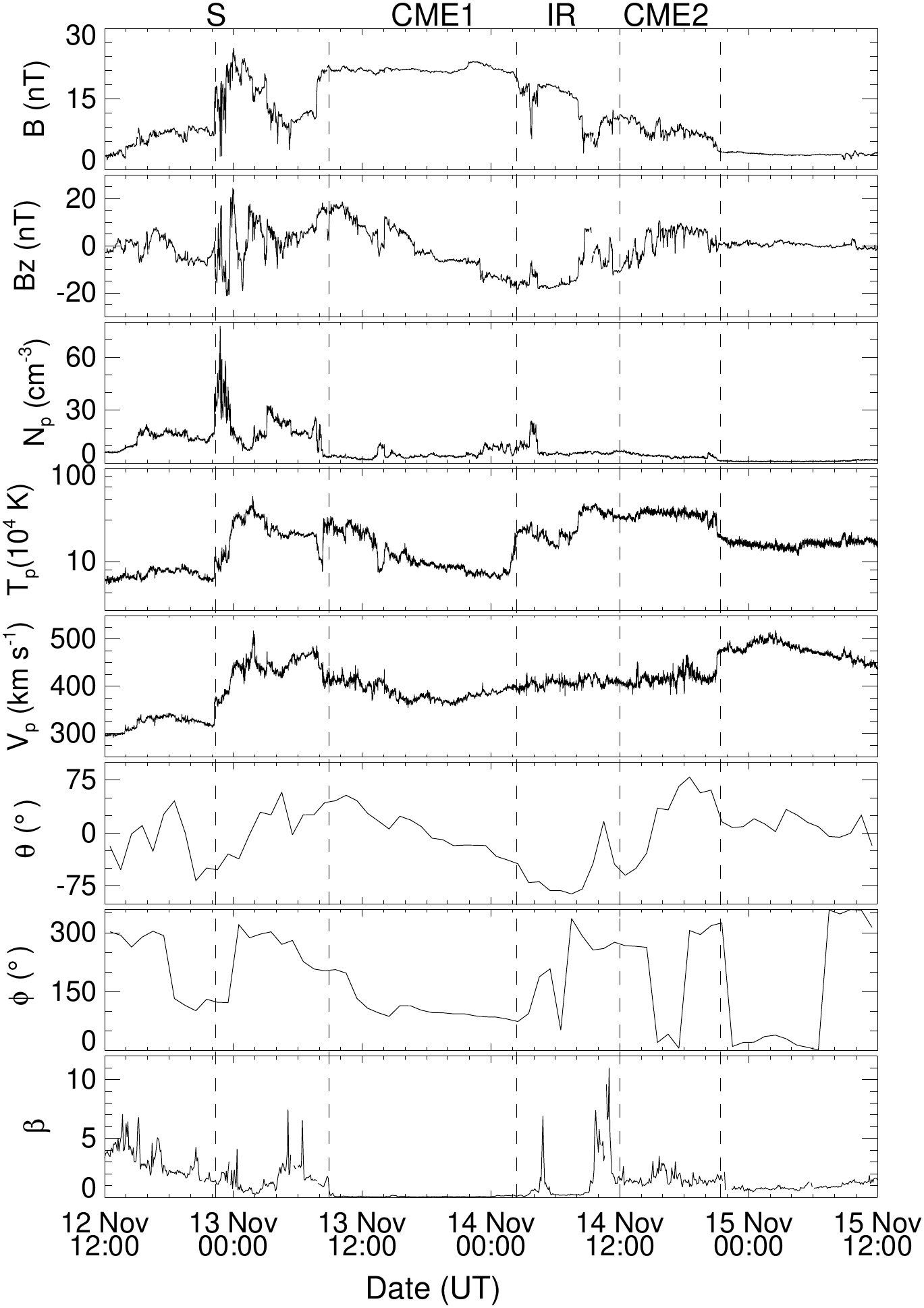}
\caption{\scriptsize{From top to bottom, total magnetic field magnitude, z-component of magnetic field, proton density, proton temperature, proton speed, latitude and longitude of magnetic field vector, and plasma beta ($\beta$) is shown for the time interval of 12:00 UT on November 12 to 12:00 UT on November 15. From the left, first, second, third, fourth and fifth vertical (dashed) lines mark the arrival of shock, CME1 leading edge (LE), CME1 trailing edge (TE), CME2 LE and CME2 TE, respectively. Interaction region (IR) is shown during the interval of third and fourth vertical lines.}}
\label{insitu}
\end{center}
\end{figure}

We analyzed the in situ data taken from WIND spacecraft located at L1 to identify the tracked density enhanced features of CMEs. Figure~\ref{insitu} shows magnetic field and plasma measurements during November 12, 12:00 UT to November 15, 12:00 UT. The arrival of a forward shock (labeled as S) marked by a sudden enhancement in speed, temperature and density is noticed at 22:20 UT on November 12. The region between the first and second vertical line represents the turbulent sheath region. Based on the CME identification criteria of \citet{Zurbuchen2006}, the region bounded between second vertical line at 08:52 UT on November 13 and third vertical line at 02:25 UT on November 14, is identified as CME structure. In this region, we observed expansion of CME characterized by a monotonic  decrease in proton speed and temperature. Based on the predicted arrival times which is derived using the estimated kinematics of the remotely observed tracked features of CME1 and CME2 as inputs in the DBM (explained in Section~\ref{arrtime}), this region is associated with Earth-directed CME1 launched on November 9. During the passage of CME1, the magnetic field is observed to be high ($\approx$ 20 nT), plasma beta is less than unity ($\beta$ $<$ 1) with smooth rotation in magnetic field vector. Also, the latitude ($\theta$) value of magnetic field vector decreased from 43$^\circ$ to -43$^\circ$ and its longitude ($\phi$) decreased from 203$^\circ$ to 74$^\circ$. Therefore, this region can be classified as a magnetic cloud (MC) and based on the observed arrival time it is associated with CME1. Due to the interaction of the CMEs, the region associated with CME1 is found to be at higher temperature than found generally for a normal isolated CME. After the third vertical line, magnetic field strength decreased reaching a minimum value of 6 nT around 4:00 UT on November 14. This interval of sudden drop in magnetic field is associated with a sudden rise in density, temperature and plasma $\beta$, suggestive of a possible magnetic hole (MH) \citep{Tsurutani2006} which is considered as a signature of magnetic reconnection \citep{Burlaga1978}. Another region of magnetic field depression from 08:05 UT to 10:15 UT on November 14, reaching a minimum value of 3 nT is also noticed. Corresponding to this minimum value of magnetic field, the plasma $\beta$ and temperature is found to  increase. The region during 03:45 UT - 08:05 UT on November 14, bounded between two distinct MH like structures, has enhanced magnetic field ($\approx$ 15 nT), and plasma $\beta$ less than unity. This region, between two MH, seems to be a magnetic field remnant of reconnecting CMEs structures. Based on the extremely elevated interval of plasma beta, temperature, observation of magnetic hole (MH) and probably sudden fast rotation in magnetic field vector during the region bounded between third and fourth vertical line, in Figure~\ref{insitu}, this region is identified as interaction region (labeled as IR) of CME1 TE with CME2 LE. Another structure is identified based on the elevated fluctuating magnetic field and temperature during 12:00 UT - 21:21 UT on November 14, bounded between fourth and fifth vertical line in Figure~\ref{insitu}. During this interval, we noticed high magnetic field (9 nT) with no monotonic decrease in temperature and speed profile as well as plasma beta ($\beta$) is not less than unity. From these observations, i.e. lack of MC signatures and the short duration (9.5 hr) of the structure associated with CME2, we infer that the WIND spacecraft perhaps intersected the flank of CME2 \citep{Mostl2010}. This is also confirmed by the estimated latitude (-25$^\circ$) of CME2 using GCS model (described in Section~\ref{3D}). On examining the evolution of CME2 in HI1-A movie, we noticed that CME2 is directed towards the southern hemisphere which is a favorable condition for its flank encounter with in situ WIND spacecraft.

We observed that the magnitude of magnetic field in MC region is constant around $\approx$ 20 nT which may be due to the passage of shock from CME1. We also noticed that average temperature in first half of CME1 is high ($\approx$ 10$^{5}$ K) than found in general ($\approx$ 10$^{4}$ K). The high temperature of the CME1 may occur due to its collision with CME2, thereby resulting in its compression. Another possibility of high temperature of CME1 is due to the passage of forward shock driven by CME2 as reported in earlier studies by \citet{Lugaz2005,Liu2012,Maricic2014,Mishra2014}. However, in the present case, the following CME2 is also observed with a high temperature ($\approx$ 5 $\times$ 10$^{5}$ K), which has not been reported in earlier studies of interacting CMEs. We noticed significant high density at the front of CME1, possibly due to overall compression by sweeping of the plasma of CME1 at its leading edge by CME2 driven shock.

\subsection{Arrival time of Tracked Features}
\label{arrtime}

We used the estimated speed at the last point of measurement (up to where CMEs could be tracked unambiguously in HI) and used it as input in the DBM developed by \citet{Vrsnak2013} to estimate the arrival time of tracked features at L1. \citet{Vrsnak2013} have shown that drag parameter lies in the range of (0.2 - 2.0) $\times$ 10$^{-7}$ km$^{-1}$. Using these values of the drag parameter in the DBM, the arrival time of CME at L1 can be predicted with a reasonably good accuracy \citep{Vrsnak2013, Mishra2013}. The estimated speed, time and distance (v$_{0}$, t$_{0}$ and R$_\odot$) of LE of CME1 (green track in J-map) are used as inputs in the DBM, corresponding to extreme range of drag parameter, to predict its arrival time and transit speed at L1. In situ observations show a peak in density, $\approx$ 0.5 hr after the shock arrival around 23:00 UT on November 12 with a transit speed of 375 km s$^{-1}$, which is expected to be the actual arrival of tracked LE feature corresponding to CME1. The predicted values of arrival time and transit speed of features and errors from the actual values are shown in Table~\ref{Tabarr}.

For TE feature of CME1 and LE of CME2, we assume that they encounter the dense ambient solar wind medium created by the preceding CME1 LE. Therefore, their kinematics with the maximum value of the statistical range of the drag parameter (2.0 $\times$ $10^{-7}$ km$^{-1}$), are used as inputs in the DBM. We consider that the TE of CME1 corresponds to the density enhancement at trailing front of CME1. At the rear edge of CME1, a density enhancement (12 particles cm$^{-3}$) is observed around 23:30 UT in WIND observations on November 13, which is considered as the actual arrival of TE of CME1. The predicted arrival time and transit speed of this feature and errors therein, are given in Table~\ref{Tabarr}. We further noticed that in situ data, an enhancement in density corresponding to LE of CME2 around 12:00 UT on November 14 (marked as arrival of CME2 with fourth vertical line in Figure~\ref{insitu}) is observed which can be considered as the actual arrival time of LE of CME2. The predicted values of CME2 LE, using its kinematics with DBM, and errors therein are also listed in Table~\ref{Tabarr}. The error for CME2 LE is large, but, several factors can lead to such large errors which are discussed in Section~\ref{Results}. It must be highlighted that if the 3D speed of CMEs estimated at final height in COR2 FOV is assumed as constant up to L1, then the predicted arrival time of CME1 and CME2 will be $\approx$ 10-16 hr and 44 hr earlier, respectively, than their predicted arrival times using post-collision speeds combined with DBM. This emphasizes the use of HI observations and post-collision speeds of CMEs as inputs in the DBM for improved arrival time prediction of interacting CMEs. Similarly, using HI observations, \citet{Colaninno2013} have shown that linear fit of deprojected height-time above 50 R$_\odot$ gives a half a day improvement over CME arrival time estimated using LASCO data.

\subsection{Geomagnetic consequences of 2012 November 9-10 CMEs}
\label{geomagnetic}

As mentioned in the introduction section, very few studies have been dedicated to the study of the geomagnetic consequences of interacting CMEs in the past. The CMEs of 2012 November 9-10 resulted in a single strong geomagnetic storm with Dst index $\approx$ -108 nT at 8:00 UT on November 14, therefore it is important to investigate the impact of the interaction of the CME1 of November 9 with the CME2 of November 10 on the terrestrial magnetosphere-ionosphere system (MI) in details.

\begin{figure}
\begin{center}
\includegraphics[width=28pc]{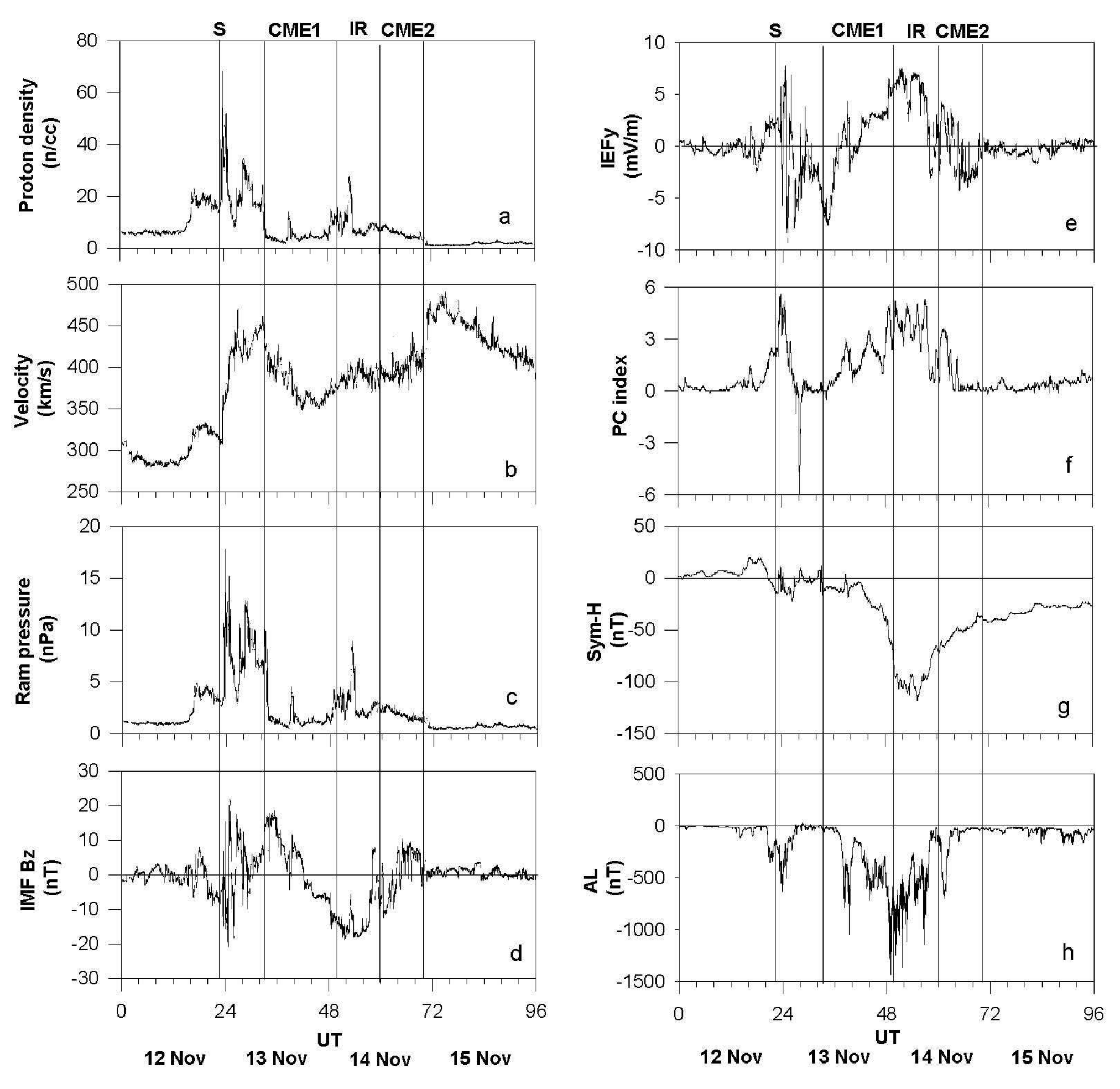}
\caption{\scriptsize{Plot shows variations of geomagnetic field parameters during the interval 2012 November 12-15. Left panels show, top to bottom: Proton density (n/cc), Velocity (km/s), Ram Pressure (nPa) and $B_{z}$ component (nT). Right panels show, from top to bottom: interplanetary electric field's 'y' component $IEF_{y}$ (mV/m), polar cap (PC) index, Sym-H (nT) and AL index (nT). From the left, vertical lines and their labels are as defined in Figure~\ref{insitu}}.}
\label{geomag}
\end{center} 
\end{figure}

Figure~\ref{geomag}(a-e) reveal the variations in the solar wind parameters in the Geocentric Solar Ecliptic (GSE) coordinate system during 2012 November 12-15. These parameters include solar wind proton density (in $cc^{-1}$), velocity (in km $s^{-1}$, negative X-direction), ram pressure (in nPa), Z-component of the interplanetary magnetic field (IMF $B_{z}$, in nT) and Y-component of the interplanetary electric field ($IEF_{y}$, in mV/m) respectively. These data with cadence of 1 minute are taken from the NASA GSFC CDAWeb (\url{www.cdaweb.gsfc.nasa.gov/istp_ public/}). It is also important to note that the solar wind parameters presented in Figure~\ref{geomag}(a-e) are corrected for propagation lag till the nose of the terrestrial bow shock. In order to compare the variations of these parameters with the magnetospheric and ionospheric parameters, additional time lags that account for the magnetosheath transit time and the Alfven transit time are calculated \citep{Chakrabarty2005}. Therefore, the solar wind parameters presented in Figure~\ref{geomag}(a-e) are corrected for the propagation lag, point by point, till ionosphere. Figure~\ref{geomag}(f) represents the variation in the polar cap (PC) index. The PC index is shown \citep{Troshichev2000} to capture the variations in the ionospheric electric field over polar region efficiently. Figure~\ref{geomag}(g) shows the variation in the Sym-H (in nT) index which primarily represents the variation in the magnetospheric ring current \citep{Iyemori1996}. The Sym-H index is essentially the high temporal resolution (1 min) version of the Dst index. Figure~\ref{geomag}(h) shows variations in westward (midnight sector) auroral electrojet (AL) current (in nT) which captures auroral substorm processes reasonably well. Further, in Figure~\ref{geomag}, the arrival of the shock is denoted by S, different vertical lines mark the arrival of different features of CME1 and CME2 and are labeled in the same manner as in Figure~\ref{insitu}.

Figure~\ref{geomag}(a) reveals the arrivals of two distinctly enhanced density structures, first enhancement occurred during arrival of the shock-sheath region before CME1 LE (on November 12 20:00 UT to November 13 04:00 UT) and second enhancement occurred during arrival of the CME1 TE-IR region (November 13 12:00 UT to November 14 12:00 UT). During the arrival of shock-sheath region before CME1 LE, the peak density reached $\approx$68 $cc^{-1}$ which is more than two times of the corresponding peak density ($\approx$30 $cc^{-1}$) observed during the arrival of CME1 TE-IR. Sharp enhancement in the solar wind velocity was also observed vis-a-vis the sharp density enhancement during arrival of shock-sheath region (Figure~\ref{geomag}b) when the velocity reached from $\approx$300 km s$^{-1}$ to $\approx$470 km s$^{-1}$. However, solar wind velocity did not change sharply and significantly during arrival of CME1 TE-IR. The changes in the density and the velocity resulted in changes in the solar wind ram pressure shown in Figure~\ref{geomag}(c). The peak ram pressure during shock-sheath region before CME LE ($\approx$18 nPa) was almost twice than that observed during CME1 TE-IR ($\approx$9 nPa). Figure~\ref{geomag}(d) reveals that IMF $B_{z}$ was predominantly southward in both density enhanced intervals although fast fluctuations were observed corresponding to the arrival of the shock-sheath region. The first peak in IMF $B_{z}$ is $\approx$-8 nT before the arrival of the shock. Thereafter, IMF $B_{z}$ fluctuated sharply between $\approx$-20 nT to $\approx$+20 nT during arrival of shock-sheath region before arrival of CME1 LE. During CME1 TE-IR, IMF $B_{z}$ reached a peak value of $\approx$-18 nT. No significant change in IMF $B_{z}$ is observed during the passage of the CME2 on November 14-15,  when its magnitude hovered around the zero line.

Figure~\ref{geomag}(e) elicits the variation in $IEF_{y}$ during 2012 November 12-15. It is observed that the peak value of $IEF_{y}$ before the arrival of the shock was $\approx$2 mV/m. However, during the shock event, $IEF_{y}$ fluctuated between $\pm$ $\sim$8 mV/m. During arrival of CME1 TE-IR, the peak value of $IEF_{y}$ reached $\approx$7.5 mV/m. In fact, similar to IMF $B_{z}$, a sharp polarity change in $IEF_{y}$ was noticed at $\sim$09:40 UT on November 14. No significant change in $IEF_{y}$ is observed during the passage of the CME2 on November 14-15, when its magnitude hovered around the zero line. Figure~\ref{geomag}(f) shows    that the PC index increased during both density enhanced intervals and the peak value is $\approx$5 for both the intervals. Figure~\ref{geomag}(g) shows variation in the Sym-H index which revealed the development of a geomagnetic storm during CME1 TE region. The Sym-H index reached a value of $\approx$-115 nT during the passage of IR on November 14. The ring current activity was not significant during arrival of the shock-sheath region before CME1. Lastly, Figure~\ref{geomag}(h) elicits the variation in the AL index. It is seen that AL reached  $\approx$-600 nT during the arrival of shock-sheath before CME1 and $\approx$-1400 nT during the arrival of CME1 TE-IR. Therefore, significant intensification of westward auroral electrojet occurred during the arrival of 
CME1 TE-IR.

The above observations (from Figure~\ref{geomag}) reveal several interesting points. First, the magnitude of AL seems to remain unaffected by large amplitudes of fluctuations in IMF $B_{z}$ and $IEF_{y}$ during the shock-sheath region. Second and the most important point is that the duration of occurrence of the AL intensification during shock-sheath before CME1 LE and in CME1 TE-IR region were nearly identical with the duration of the southward IMF $B_{z}$ and positive $IEF_{y}$ phases. However, the AL amplitudes seem to be more closely correlated with the $IEF_{y}$ amplitudes during CME1 TE-IR region compared to shock-sheath region. Third, although the peak amplitudes of PC index were nearly same during shock-sheath region before CME1 LE and CME1 TE-IR region, the peak amplitudes of the AL index were significantly different during these two intervals. Fourth, the substorm activity seems to be over during the passage of the CME2, therefore, CME2 did not have any bearing on the triggering of substorms. Figure~\ref{geomag} demonstrates the direct role played by the positive $IEF_{y}$ (or southward IMF $B_{z}$) in the storm-time AL intensification particularly when the terrestrial magnetosphere encounters the CME1 TE and IR region. Further, the arrival of CME2 did not affect the terrestrial magnetosphere-ionosphere system.

As aforementioned, major geomagnetic response is noticed during the arrival of trailing edge of the preceding CME (CME1) and the interaction region (IR) of the two CMEs near the Earth. We also conclude that the following CME (CME2) failed to cause a significant geomagnetic activity possibly due to the fact that the in situ spacecraft encountered the flank of this CME as mentioned in Section~\ref{Insitu}. We understand that due to interaction and collision between the trailing edge of the preceding CME1  and the leading edge of the following CME2 as revealed in imaging observations (described in Section~\ref{HI}), the parameters responsible for geomagnetic activity were significantly intensified at CME1 rear edge and in interaction region found between CME1 and CME2. These results bring out the importance of CME-CME interaction in the formation of interaction region and its role in the significant development of geomagnetic disturbances.

\section{Results and Discussion}
\label{Results}

In this paper, we have studied the interactions between two successively launched Earth directed CMEs from the Sun on 2012 November 9 and 10, respectively. The analysis involved remote sensing observations of CME propagation in inner heliosphere from SECCHI/COR2 and HI images and the associated in situ measurements from WIND spacecraft. The first CME (CME1) associated with a filament eruption is observed to be slower (600 km s$^{-1}$) than the successive CME (CME2) launched on November 10 with a higher velocity (900 km s$^{-1}$). Both the CMEs were launched in approximately the same direction towards the Earth, indicating a possibility of interaction because of their relative speeds and the same direction. Different features were tracked as brightness enhancements in J-maps obtained from COR2 and HI images and subsequently HM technique was applied to estimate the kinematics of the tracked features of the CMEs. From the estimated kinematics, the site of collision of CMEs could be located at a distance of approximately 35 R$_{\odot}$ at 12:00 UT on November 10, which is at least 85 R$_{\odot}$ before and 15 hours earlier than as predicted by using the initial kinematics of CMEs in COR2 FOV. The estimated kinematics were then used as inputs in the DBM to determine the arrival times of the tracked  features. After the interaction, CME2 transferred some of its velocity to CME1 and  both were observed to propagate together, showing a distinct deceleration of CME2.

As described earlier, CME1 LE is found to be propagating with higher speed than CME1 TE. Therefore, CME1 LE (green track) did not collide with CME2 LE (blue track). With the start of collision phase, the LE of CME1 is also found to accelerate and after the collision phase, significant acceleration of CME1 LE is noticed. This may occur either because of sudden impact (push) from CME1 TE to CME1 LE during collision or due to the passage of a shock driven by the CME2 or a combined effect of both. Beyond the collision phase, up to an estimated distance of nearly 100 R$_{\odot}$, LE of CME2 was found moving behind the TE of CME1 and both these features propagated together decelerating slowly. For few hours after the observed collision phase, it is noticed that speed of the LE and TE of CME1 is slightly higher than LE of CME2, which can increase the observed separation of these structures. The 
J-maps show that both these features can be tracked for further elongations after interaction. However, due to the limitation of the HM method for higher elongations \citep{Lugaz2009}, we restricted our measurements on the tracked features up to 100 R$_{\odot}$. Our analyses show an evidence that after the interaction, the two features did not merge. It appears that LE of CME2 interacted with TE of CME1 and continued to propagate with a reduced speed of 500 km s$^{-1}$. In the SECCHI/HI-B images these features cannot be well distinguished and therefore it is difficult to use any stereoscopic method to infer about a possible longitudinal deflection (based on stereoscopic methods) of these features after interaction.

During the collision between CME1 and CME2, we noticed a large deceleration of fast CME2 while relatively less acceleration of slow CME1, till both approached an equal speed. This is expected to occur if the mass of CME1 is larger than that of CME2, which is indeed the case. In Section~\ref{Momentum} we have shown that the mass of CME1 is $\approx$ 2.0 times larger than the mass of CME2. This result is an important finding and is in agreement with the second scenario of interaction described in \citet{Lugaz2009}. It also must be noted that we have estimated the mass of CMEs in COR2 FOV while interaction takes place in HI FOV. Further, we cannot ignore the possibility of increase in mass of CME due to mass accretion at its front via snowplough effect in the solar wind beyond COR FOV \citep{DeForest2013}. We must also mention that in our calculation of momentum and energy transfer, although we use the total mass of CME1 but only a part of CME1 (i.e. TE of CME1) takes part in collision with CME2. Considering the uncertainties in the derived speeds and mass of the CMEs, our analysis reveals that the nature of collision remains close to perfectly inelastic. However, we also acknowledge that defining collision phase is crucial and seems to play an important role in this calculation. In this context, it must be emphasized that marking of the start of collision is often difficult. This is because the following fast speed CME2 starts to decelerate, well before the actual merging or collision is revealed in imaging observations. This is also shown in earlier studies for other interacting CMEs \citep{Temmer2012,Shen2012,Maricic2014,Mishra2014}. The interaction of CME2 with streamers or overlying coronal magnetic field lines \citep{Temmer2008} may be responsible for deceleration of following CME2 before its collision with CME1 takes place. However, the contribution of other factors, e.g., the solar wind drag acting on CME2, interaction of CME2 with trailing edge of preceding CME1, and magnetic interaction of CME2 with preceding CME1 cannot be completely ignored. Also, possibly a shock launched by the following CME2 can contribute in acceleration of the preceding CME1 before the actual merging of both CMEs. Therefore, different timing and large time-interval of acceleration of one CME and deceleration of other, prevent us to pinpoint the exact start and end of the collision phase.

We concluded that the collision of interacting CMEs of 2012 November 9-10 is close to perfectly inelastic in nature. Here, it is noted that we are dealing with large scale magnetized three-dimensional structures, however, in the present study we have reduced the calculations to one-dimensional collision which seems to be accurate as tracked features of both CME1 and CME2 are considered to move along approximately the same trajectory. This implies that there will be a small difference between velocity of tracked feature of one CME and its component along the direction of propagation (i.e. longitude) of tracked feature of other CME. We must admit that possibility of change in longitude of tracked features during their collision can partially contribute to some errors in the estimated speed. In light of earlier reported observations \citep{Shen2012,Temmer2012,Mishra2014}, it seems that collision of CMEs can occur in all i.e. elastic, inelastic or super-elastic regimes. An in-depth study is required further for understanding the nature of CME-CME collision which may depend on the characteristics of CMEs, locations and duration of collision phase in the heliosphere.

At the last point of measurement in HI FOV around 120 R$_{\odot}$, the estimated speed of TE of CME1 and LE of CME2 are approximately equal ($\approx$ 470 km s$^{-1}$). Therefore, both features are expected to arrive at L1 at approximately the same time. Comparing to in situ measured actual arrival time, the delayed arrival of LE of CME2 is possible due to higher drag force acting on it resulting in its deceleration. However, the propagation direction of CME2 in southward hemisphere, as noticed in HI1-A images and also estimated in COR2 FOV using 3D reconstruction, can also account for flank encounter of CME2 and thus its delayed sampling by WIND spacecraft at L1. This is consistent with our interpretation in Section~\ref{Insitu}. Since, in the HM method, the front of the CME is assumed spherical which may not be the real case for CME2, as it was moving with fast speed and has large probability to distort and flatten its front while interacting with structured solar wind \citep{Odstrcil2005}. In this case, the estimated speed using HM method will lead to some error and can result in the delayed arrival of LE of CME2. Keeping all these issues in mind, we believe that it is quite probable that the tracked feature corresponding to LE of CME2 (blue track in J-map) is not sampled by in situ spacecraft. Therefore, erroneous predicted arrival time for LE of CME2 is related to incorrect identification of tracked CME2 LE in in situ data which is taken as a reference (actual) for remotely observed tracked CME2 LE.

The present study not only estimates the kinematics and arrival times based on HI observations and their application to the DBM model of CME propagation, it also attempts to associate tracked brightness enhancements as observed in remote sensing COR and HI (J-maps) with the in situ observations \citep{Mostl2010,Mostl2011,Rollett2011,Rollett2013}. The association of HI observations with in-situ measurements from WIND leads to many interesting results. In spite of two CMEs launched from the Sun in succession in the Earthward direction, we observe only one shock in in situ data which may suggest the merging of shocks driven by CME1 and CME2, if both CMEs would have driven shocks. However, such a claim cannot be made unless we are well familiar with the in situ signatures of merged shock and plasma structure following it. The sweeping of plasma to high density at the front of CME1 and its compressed heating is most likely due to passage of CME2 driven shock through the MC associated with CME1. Based on the predicted arrival time of tracked features, it seems that the CME2 driven shock and CME1-sheath region is tracked as CME1 LE in J-map. Therefore, we infer that CME1 LE is propagating probably into an unperturbed solar wind. Our study also provides a possibility of formation of interaction region (IR) at the junction of trailing edge of preceding CME and leading edge of following CME. We show that during collision of the CMEs, kinetic energy exchange up to 50\% and momentum exchange between 23 to 30\%. Our study also demonstrate that the arrival time prediction significantly improved using HIs on-board STEREO compared to COR2 observations, and also emphasizes the importance of understanding of post-collision kinematics in further improving the arrival time prediction for a reliable space weather prediction scheme.

Our study reveals clear signatures of interaction of these CMEs in remote and in situ observations and also helps in identification of separate structures corresponding to these CMEs. In spite of the interaction of the two CMEs in the interplanetary medium which generally results in complex structures as suggested by \citet{Burlaga1987,Burlaga2002}, in our case, we could identify interacted CMEs as distinguished structures in WIND spacecraft data. Even after collision of these CMEs, they did not merge which may be possibly because of strong magnetic field and higher density of CME1 than CME2. This needs further confirmation, and therefore, it is worth to investigate what decides the formation of merged CME structure or complex ejecta during CME-CME interaction. Due to single point in situ observations of CME, we acknowledge the possibility of ambiguity in marking the boundaries of CMEs. In the present case, the boundaries for CME1 and IR are distinctly clear. Also, slight ambiguity in the boundary of CME2 (flank encounter) will not change our interpretation because the main geomagnetic response is caused due to enhanced negative Bz in trailing portion of CME1 and its extension in IR. Here, we also point out that temperature in IR is lesser than the temperature for CME2 region but temperatures in both regions are elevated as compared to a normal non-interacting CME. The observations of unexpected larger temperature in CME2 region than IR region may be due to the possibility that sheath region of CME2 is intersected by in situ spacecraft, as we have interpreted.

The association of geomagnetic storms with isolated single CMEs has been carried out extensively for a long time \citep{Tsurutani1988,Gosling1991,Gosling1993,Gonzalez1989,Gonzalez1994,Echer2008,Richardson2011}. However, only few studies have been dedicated to understand the role of interacting CMEs in the generation of geomagnetic storms \citep{Burlaga1987,Farrugia2006,Wang2003,Xiong2006}. Our study is important as it focuses on the role of interacting CMEs in the generation of geomagnetic storms as well as substorms. In the context of substorms, our study highlights that persistence of IMF Bz in the southward direction is more important than the amplitude in driving the substorm activity as manifested by the AL intensification. Using WINDMI model, \citet{Mays2007} have shown that the IP shock and sheath features for CMEs contribute significantly to the development of storms and substorms. But, in our study, sharp and large southward excursions in the midst of fluctuating IMF Bz associated with shock (shock-sheath region before CME1) were found less effective in producing strong substorm activity. 
Therefore, further investigations are required regarding characteristics (geometry, intensity) of shock and preceding CME in the context of triggering of substorms, as has been shown in earlier studies \citep{Jurac2002,Wang2003a}. Regarding geomagnetic storms, our study suggests that the trailing edge of preceding CME (CME1) and IR formed between two interacting CMEs are efficient candidates for intense geomagnetic storms.

\setlength{\emergencystretch}{1em} {Another interesting aspect regarding substorms noticed in our study} is that the nearly equal amplitude responses of the PC index corresponding to the shock-sheath region (IMF Bz sharply fluctuating between southward and northward directions) preceding the CME1 interval vis-\`a-vis the CME1 interval (IMF Bz steadily turning southward). This is interesting as the responses of Sym-H and AL during these two intervals are quite different in terms of amplitudes of variations. \citet{Janzhura2007} have inferred that the PC index for the sunlit polar cap (summer hemisphere, high polar ionospheric conductance) responds mainly to the geo-effective interplanetary electric field whereas the PC index for the dark winter cap (winter hemisphere, low polar ionospheric conductance) responds better to the particle precipitation in the auroral zone like the AE and AL indices. It is to be noted here that, in the present study, no distinction is made between the variations of PC index in the northern (PCN) and southern (PCS) hemispheres. On the other hand, it is known that AL index is constructed based on magnetometer observations around the northern auroral oval only. The discrepancy between the variations in the PC index and the AL index may, therefore, suggest towards the hemispherical asymmetry in the response of polar ionosphere corresponding to this event. This is not unexpected as asymmetric auroral intensities in the northern and southern hemispheres are reported in literature \citep{Laundal2009}. Further investigations are needed to understand this aspect.

\section{Conclusions}
\label{Conclu}

The CMEs of November 9 and 10, provide us a rare opportunity to investigate the consequences of CME-CME interaction. A combination of heliospheric imaging and in situ observations are used for improving our understanding of CME kinematics, post-collision characteristics and nature of collision. Main results of our study can be concluded as follows:

\begin{enumerate}

\item{The analysis of propagation kinematics as obtained from J-maps provide evidence that the CME1 and CME2 collide at 35 R$_{\odot}$ much earlier than by using the estimated kinematics in the COR2 FOV. This emphasizes the importance of heliospheric imaging particularly, for interacting CMEs and ascertaining their impact and arrival at the Earth.}

\item{Our analysis shows that post interaction and collision kinematics is required for a better prediction of CME arrival time at 1 AU. In the present case of November 9 and 10 CMEs, the speeds and momentum of CMEs changed from 23\% to 30\% compared to their values before the collision. Our results also highlight that estimated kinematics, in particular after collisions are important to combine with DBM for improving the estimation of arrival times of different features of CME which experience different drag forces during their propagation through the heliosphere. Based on this study, we conclude that CMEs cannot be treated as completely isolated magnetized plasma blobs, especially when they are launched in quick succession.} 

\item{Using estimated mass and kinematics before and after the collision, we estimated the total kinetic energy of the system before and after the collision. We found that total kinetic energy of the system decreased by 6.7\% to its value before the collision and the nature of collision is close to perfectly inelastic. The study also supports the idea that CME interaction or collision can lead to the heating and compression of both preceding and following CMEs.}

\item{It is clear that if the trailing edge of November 9 CME would not have been tracked using imaging observation, one could not have witnessed the collision. Further, trailing edge of this CME had strong negative component of magnetic field for long duration ($\approx$ 13 hr) which may be due to its collision with the leading edge of November 10 CME. Therefore, our study reveals that tracking of different features of CME seems to be necessary for better understanding of CME-CME interaction.} 

\item{It is observed in the present case that the persistence of IMF $B_{z}$ in the southward direction is more important (rather than the amplitude) in driving the substorm activity as manifested in the AL intensification. Sharp and equally large southward excursions in the midst of fluctuating IMF $B_{z}$ associated with shock were found less effective in producing equally strong substorm activity. It is identified that the interaction region (IR), formed due to collision between CME1 TE and CME2 LE, has intensified plasma and magnetic field parameters which are responsible for major geomagnetic activity.}

\end{enumerate}
The present case study of interacting CMEs of 2012 November 9-10, highlight the importance of heliospheric imaging for estimating the kinematics of CME features before and after their collision and interactions. Our study support that interacting CMEs can result in strong geomagnetic storms and also in substorms. Further statistical study of such interacting CMEs is required to understand their nature of collision and to investigate the characteristics of the CMEs (mass, strength and orientation of magnetic field, speed and direction of propagation, duration of collision phase) which are responsible for their interaction.

\clearpage

\begin{acks}
We acknowledge the UK Solar System Data Centre for providing the processed Level-2 STEREO/HI data.  The in situ measurements of solar wind data from ACE and WIND spacecraft were obtained from NASA CDAweb (\url{http://cdaweb.gsfc.nasa.gov/}). We also acknowledge the use of DBM, developed by Bojan Vr$\breve{s}$nak and available at \url{http://oh.geof.unizg.hr/CADBM/cadbm.php}, in this study. We also thank the reviewer for useful comments, which helped improve the manuscript. The work by N.S. partially contributes to the research for European Union Seventh Framework Programme (FP7/2007-2013) for the Coronal Mass Ejections and Solar Energetic Particles (COMESEP) project under Grant Agreement No. 263252.
\end{acks}

\clearpage

\pagestyle{empty}
\begin{landscape}
\begin{table}[htbp]
  \centering
{\scriptsize
 \begin{tabular}{ p{1.5cm}p{2.7cm} p{3.7cm} p{3.0cm}p{3.0cm}p{3.2cm}}
    \hline
		
 Tracked Features & Kinematics as inputs in DBM [t$_{0}$, R$_{0}$ (R$_\odot$), v$_{0}$ (km s$^{-1})$]& Predicted arrival time (UT) using kinematics + DBM [$\gamma$ = 0.2 - 2.0 (10$^{-7}$ km$^{-1}$)] & Predicted transit speed (km s$^{-1}$) at L1   [$\gamma$ = 0.2 - 2.0 (10$^{-7}$ km$^{-1}$)] & Error in predicted arrival time (hr) [$\gamma$ = 0.2 - 2.0 (10$^{-7}$ km$^{-1}$)] &  Error in predicted speed (km s$^{-1}$)  [$\gamma$ = 0.2 - 2.0 (10$^{-7}$ km$^{-1}$)]  \\  \hline

CME1 LE & Nov 11 13:42, 545, 137  & Nov 12 18:10 - Nov 13 00:30  & 490 to 380 	& -5 to 1.5 	& 115 to 5  \\ \hline

CME1 TE & Nov 11 18:35, 120, 470  & Nov 13 15:25                  & 375        &  -8         &  -15  \\  \hline

CME2 LE & Nov 12 00:30, 124, 455  & Nov 13 19:40                 	& 375   	   &  -16        &  -35  \\ \hline

\end{tabular}
}
\caption{\scriptsize{First column shows the tracked features of CMEs and second column shows their estimated kinematics (by HM technique) which is used as input to the DBM. The predicted  arrival time  and transit speed of the tracked features at  L1, corresponding  to the extreme range of the drag parameter used in the DBM, is shown in the third and fourth  column. Errors in predicted arrival time and speed, based on comparison with in situ arrival time and speeds, are shown in column fifth and sixth. The errors in arrival time with negative (positive) sign indicate that predicted arrival is earlier (later) than the actual arrival time. The errors in transit speed with negative (positive) sign indicate that the predicted transit speed is lesser (greater) than the transit speed measured in situ.}}
\label{Tabarr}
\end{table}

\end{landscape}

\pagestyle{plain}

\clearpage

\clearpage


\end{article} 
\end{document}